# Topological quantum materials: kagome, chiral, and square-net frameworks

Avdhesh K. Sharma, Snehashish Chatterjee[#], Premakumar Yanda[#], Claudia Felser[#] and Chandra Shekhar[#]

Max Planck Institute for Chemical Physics of Solids, 01187 Dresden, Germany

**Abstract**

Topological quantum materials have emerged as a frontier in condensed matter physics as well as in materials science, with intriguing electronic states that are robust to perturbations. Among the diverse structural motifs, kagome, chiral, and square-net structures offer a wide range of topological phases and physical phenomena. These include Dirac and Weyl fermions, nodal-line semimetals, flat bands, van Hove singularities, charge density waves, superconductivity, nontrivial Berry phase, nonlinear electrical and thermal transports. This review explores the distinct roles of geometry, symmetry, spin-orbit coupling, and electron correlations in these three classes of materials. It also highlights how their crystallographic features give rise to unique electronic band structures, topologically protected states and different physical properties, which require high-quality-single crystals. The present discussion comprises recent experimental discoveries and identification of the synthesis routes of key materials within each framework. Finally, the review outlines the current challenges and future directions in the design and exploration of topological quantum materials.

---

[#]email: chatterjee@cpfs.mpg.de; yanda@cpfs.mpg.de; felser@cpfs.mpg.de; shekhar@cpfs.mpg.de

**Introduction**

Topological quantum materials are a class of condensed matter systems whose properties are governed by strong quantum mechanical effects, often emerging from the interplay of spin, charge, orbital, and lattice degrees of freedom. These characteristics give rise to emergent phenomena, including unconventional superconductivity, topological behavior and exotic quantum spin liquids[1-4]. In recent years, an expanding array of quantum materials has been discovered, many of which are distinguished by the geometry of their underlying crystal lattices. Notably, materials featuring kagome, chiral, and square-net lattices have become central to the exploration of quantum phases of matter because of their unique symmetry properties and topological characteristics, which give rise to a variety of quantum properties (Fig. 1)[5-11]. Among these, the kagome lattice has garnered particular interest for hosting a combination of nontrivial electronic features such as the Dirac cone (DC)[8-10], flat band (FB)[8-10], and van Hove singularities (vHS)[7-10]. The vHS typically appears at the zone edge $M$ of the Brillouin zone (BZ), while the inclusion of spin-orbit coupling (SOC), both the quadratic band touching point at the center $\Gamma$-point and the linear band crossing at the corner $K$ are gapped[12,13]. These phenomena are interconnected with magnetism and SOC and lead to several quantum physical properties, including unconventional charge density wave (CDW)[13-18], unconventional superconductivity[13-18], and intrinsic anomalous Hall effect (AHE)[19-22]. The hybridization strength and the type of atoms located at the corners of the kagome lattice influence the extent of distortion experienced by the lattice. Such distortions may be attributed to atomic shifts, chemical doping, or anisotropies in the crystal lattice[23]. These factors can result in alterations in bond lengths and angles, thereby reducing the inherent geometrical frustration of the perfect kagome structure. These systems are referred to as distorted kagome structure and exhibit diverse quantum states, including spin ice, Weyl fermions, and superconductivity[24-26]. Another distinct family of quantum materials is chiral materials, which lack inversion symmetry and often crystallize in enantiomorphic space groups. The chiral crystals allow a novel type of topologically protected massless fermions, termed as Kramers-Weyl fermions linked with giant arc-like surface states, which emerge at high symmetry points in the BZ and are protected by chiral symmetries[5,27-31]. The protection at high symmetry points allows for three-, four-, or six-fold crossings (multifold fermions) associated with higher Chern numbers and higher-order fermions[28]. The absence of inversion symmetry further enables exceptional phenomena, such as nonlinear transports, and the chiral arrangement of atoms provides an additional parameter to observe electronic magnetochiral anisotropy[32]. In magnetic chiral materials, the Dzyaloshinskii-Moriya interaction stabilizes unconventional arrangements of spins into skyrmions and describes how the SOC in a crystal lattice can lead to a preferential twisting or canting of magnetic spins[33]. In contrast, square-net materials are composed of atoms arranged in planar $4^4$ nets, which enable the realization of nodal-line semimetallic states and Dirac-like crossings near the Fermi energy ($E_F$)[11,34]. These materials crystallize in nonsymmorphic space groups, where glide-mirror or screw-axis symmetries protect band degeneracies across specific high-symmetry lines in the BZ[35,36]. This leads to the formation of drumhead surface states, a hallmark of nodal-line semimetals. While certain Dirac-type crossings arising from square-net $p_x/p_y$ orbital overlaps are symmetry-protected, others are accidental and become gapped in the presence of SOC, particularly when inversion or mirror symmetries are broken[11,35,36]. Moreover, in magnetic square-net systems, the combined effect of SOC and broken time-reversal symmetry gives rise to Weyl nodes, a large Berry curvature, and associated topological transport phenomena, such as the AHE and anomalous Nernst effect (ANE). The unique electronic structure driven by the square-net geometry, which can be tuned through magnetism or chemical substitution, has established these materials as an emerging platform for exploring symmetry-driven and interaction-enhanced topological quantum states.

This review provides a comprehensive overview of the synthesis and quantum phenomena in diverse series of kagome, chiral, and square-net quantum materials, which are emerging materials that host exotic electronic states and topological phases. The synthesis of these quantum materials presents significant challenges. Achieving high crystal quality and controlled stoichiometry is imperative because minor deviations can significantly alter their electronic structures. Various synthesis techniques have been explored, focusing on the physical and thermodynamic characteristics of these materials.

**Tool box for quantum materials and phenomena**

- **Anomalous Hall effect** is defined as the occurrence of a non-zero Hall effect in a zero-external magnetic field. This effect arises from the interplay between spin-orbit coupling and the magnetic properties of the material, specifically the presence of broken time-reversal symmetry.

- **Anomalous Nernst effect** is the thermal counterpart of the anomalous Hall effect.

- **Berry phase** is a geometric phase acquired by an electron when it moves in a closed loop.

- **Charge density wave** is a state of matter in which electrons form a periodic, wave-like modulation of their density, accompanied by a distortion of the underlying crystal lattice. Therefore, the distribution of electrons is not uniform, but rather accumulative in certain regions and a scarcity in others.

- **Chirality** refers to the property of an object that distinguishes it from its mirror image, and it cannot be superimposed onto that image as one would with left and right hands.

- **Dirac semimetal** is a distinct quantum material where electrons behave as massless particles, exhibiting a linear energy dispersion near the Fermi energy, similar to graphene. These materials are considered 3D analogs of graphene.

- **Dzyaloshinskii-Moriya interaction** is an antisymmetric exchange interaction between neighboring spins in a magnetic system. This phenomenon emerges in systems that exhibit a lack of inversion symmetry and have strong spin-orbit coupling.

- **Electrical magnetochiral anisotropy** is a nonlinear, directional, and magnetic-field-dependent transport phenomenon, which arises due to chiral nature of the system. It means the electrical resistance depends on both the direction of current flow, the applied magnetic field, and chirality.

- **Flat band** refers to electronic band structure in which the energy of the electrons does not change with momentum, that is, it is dispersionless. This can occur under specific conditions, such as strong electronic correlations, specific lattice geometry, or the presence of an external magnetic field.

- **Pair density wave** is a pattern of Cooper pairs that are not evenly spread out in a material. Instead, they form a repeating pattern that resembles a wave.

- **Quantum spin liquid** is a state in which electron spins remain disordered and fluctuating even at the lowest possible temperature. Contrary to the characteristic behavior of conventional magnetic materials, which exhibit ordered spins at a specific temperature.

- **Topological Hall effect** is a type of the Hall effect that arises due to the real space Berry phase acquired by electrons. This phase is a consequence of the existence of noncoplanar spins, such as skyrmions.

- **Topological insulator** exhibits insulating behavior in its bulk but surface is conducting. The surface states are characterized by a non-trivial electronic band structure, which provides a degree of protection.

- **Superconductor** conducts electricity without energy dissipation below a critical temperature. A quantum mechanical pairing of electrons called Cooper pairs form due to an attractive interaction between electrons thus eliminating electrical resistance.

- **Skyrmion** is a swirling structure of spins and a topological object, meaning it possesses a kind of twist or winding that cannot be undone without breaking the structure, much like a knot.

- **Spin ice** is a state of a magnetic material where the magnetic spins are highly frustrated, leading to an ice-like arrangement even at very low temperatures.

- **van Hove singularity** is defined by a divergence or a sharp increase in the density of states, which is attributed to the presence of a saddle point within the band structure.

- **Weyl semimetal** is a class of materials in which conduction and valence bands touch at specific points in momentum space, called Weyl nodes. These points are in opposite chirality and are further connected with Fermi arc.

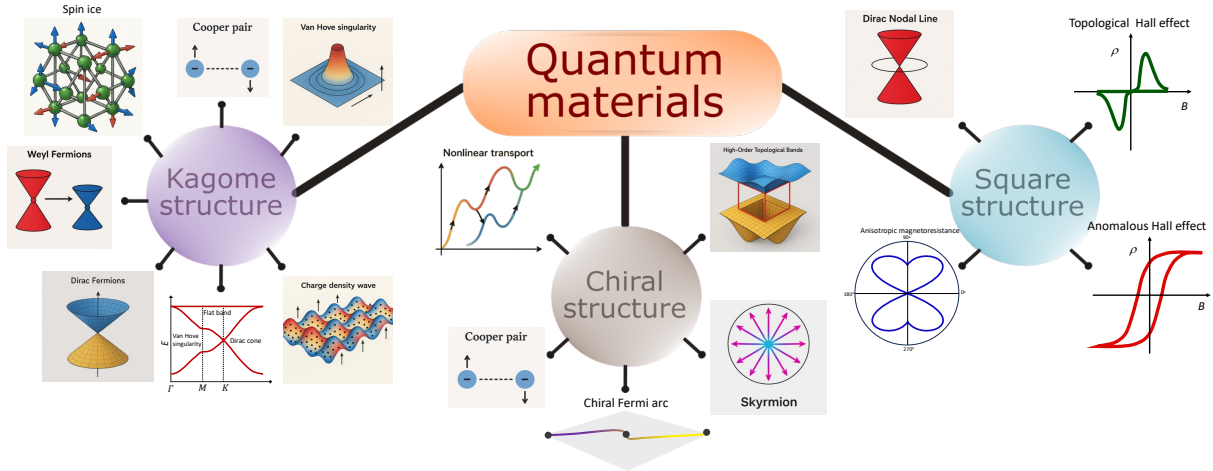

**Fig.1 Key physical quantum phenomena.** Various physical properties of kagome, chiral and square-net structured materials.

## 1. Kagome Materials

Kagome materials are a class of crystalline solids in which atoms are arranged in a two-dimensional network of corner-sharing triangles, resembling the kagome lattice known from traditional Japanese basket-weaving patterns[37]. Figure 2 illustrates the structure of several series of kagome materials, demonstrating both along the *c*-axis and perpendicular to it. The distinctive geometry leads to strong geometric frustration, resulting in key electronic features such as DC, FB, and vHS [7-10]. These properties make kagome materials ideal platforms for exploring a wide range of correlated and topological quantum phenomena. These exotic states include quantum spin liquids, CDWs, Dirac fermions, and unconventional superconductivity[13-18]. These remarkable properties position these materials as a significant series in the expansive realm of quantum materials. In this section, we discuss the different series of kagome materials and their emergent key properties, which explain the need for synthesizing more kagome materials to advance the understanding of quantum materials[1-4].

### 1.1 $M_mX_n$

Topological research on kagome materials started with this series and has achieved several milestones in the timeline of the discovery of topological quantum phenomena. These milestones include the observation of Berry phase-induced AHE, FB, and massive Dirac gap[19,20,38,39]. The general composition of the series is $M_mX_n$ (where; *M* is Co, Mn, Fe; *X* is Sn, Ge; and m, n = 1-3). Notably, most magnetic compounds exhibit magnetic transitions at temperatures above room temperature, suggesting potential applications in various fields[40-43]. The series under consideration includes $Mn_3Ge$, $Mn_3Sn$, $Fe_3Sn_2$, FeSn, and FeGe. The non-collinear inverse triangular spin order of $Mn_3Ge$ and $Mn_3Sn$ forms a chiral spin structure, which has been identified as the source of the Berry phase and has been found to give rise to AHE and ANE even at room temperature (Fig. 3a)[19-22]. $Fe_3Sn_2$ is a bilayer kagome, and ferromagnetism induces a large band gap in the Dirac bands[12,38,44]. In contrast, FeSn is a collinear antiferromagnetic (AFM) wherein Fe atoms form a distorted body-centered cubic arrangement with Sn atoms, and CoSn is a Pauli paramagnet, exhibiting Dirac fermions and FBs, as demonstrated in experimental studies (Fig. 3e)[39,45]. FeGe is distinct from other compounds in the series, displaying AFM behavior below 400 K and a CDW at $T_{CDW}$ ~110 K[46-48]. This is the result of involving a multifaceted interplay of electron correlations, magnetic order, vHS, and CDW. The CDW in FeGe is characterized by a multiple-$q$ modulation, with wavevectors connecting the vHS at high-symmetry points in the BZ. This modulation is accompanied by the partial dimerization of Ge atoms along the $c$-axis, leading to a periodic lattice distortion. Notably, the CDW does not emerge from the conventional phonon softening mechanism. Rather, it is driven by spin-phonon coupling and magnetic exchange interactions, highlighting its unconventional origin[48-50].

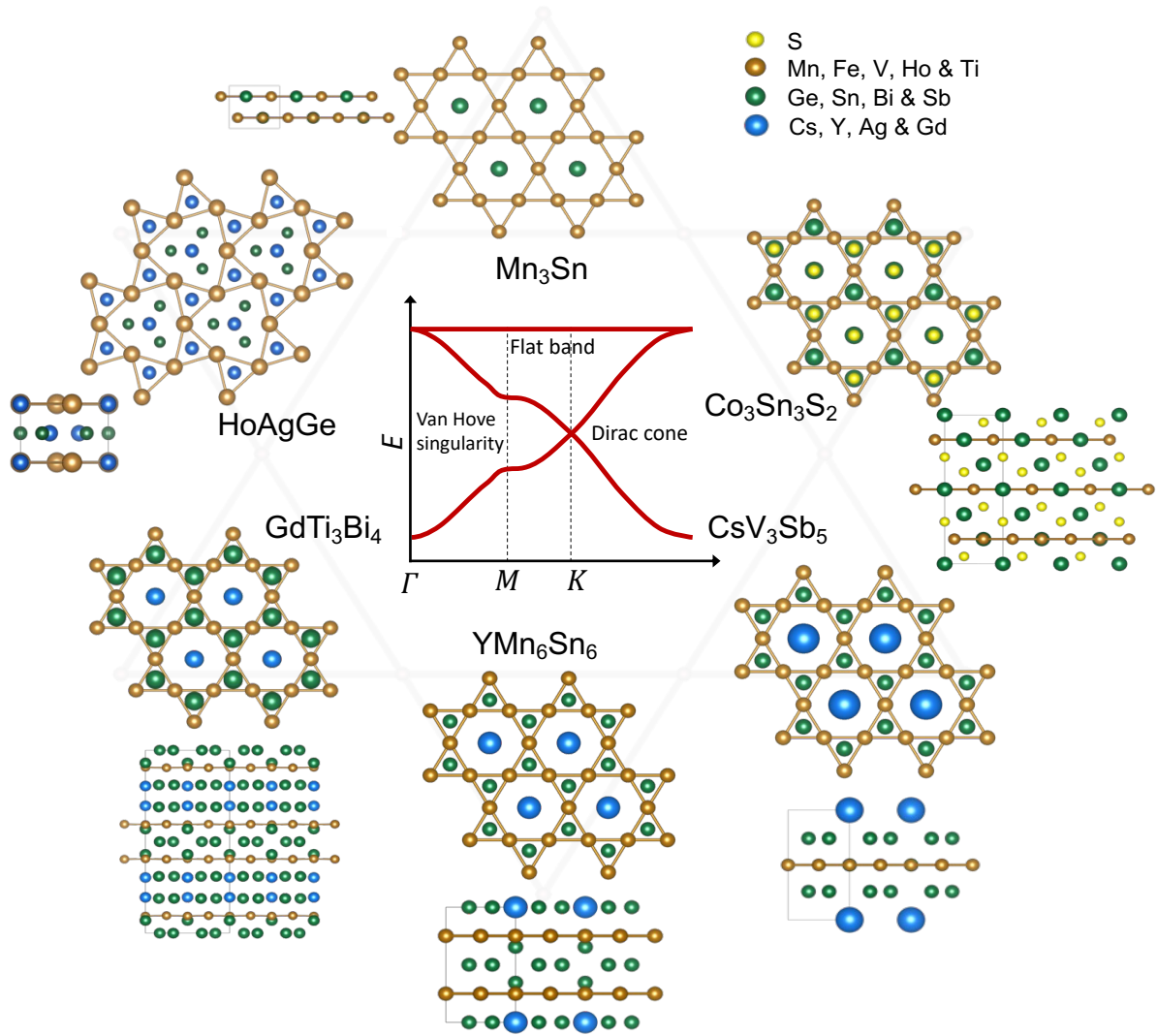
6

**Fig. 2 Structural arrangement of different kagome materials.** The unit cell of each kagome family compound hosts a distinct arrangement of atoms, giving rise to benchmark electronic features of Dirac cones, flat bands, and van Hove singularities, which are hallmark properties of this class of quantum materials. The structures are illustrated along the *c*-axis and perpendicular to it.

### 1.2 Shandite Series

This series of materials is generally represented by the formula $M_3X_2X'_2$ (where $M$ = Co, Ni, Pd, or Rh; $X$ = Sn, In, Sb, Pb, or Bi; and $X'$ = S, Se, or P). These materials possess a quasi-two-dimensional layered structure characterized by stacked kagome layers[51]. Among them, $Co_3Sn_2S_2$ stands out as the only ferromagnetic compound, exhibiting a half-metallic band structure driven by strong magnetic anisotropy with out-of-plane magnetization[52]. The kagome network is formed by Co atoms, whereas the triangular lattices are composed of Sn and S atoms (Fig. 2). This unique structure supports a large anomalous Hall conductivity (AHC) of ~1100 $\Omega^{-1}cm^{-1}$ and an anomalous Hall angle of ~20%, both of which remain constant up to 100 K[52]. Notably, this temperature range is exceptionally uncommon, with a $T_C$ of ~ 175 K. The electronic bands form a nodal ring, which partially opens the gap upon the inclusion of SOC, thereby creating a pair of Weyl nodes close to the $E_F$. This results in a non-zero Berry curvature, which is the source of the large observed AHE. Remarkably, the quantized contribution of AHC per magnetic kagome layer reaches $e^2/h$, highlighting the intrinsic topological nature of the system (Fig. 3b)[4]. These are supported by the presence of step-like 1D chiral edge states at the terrace of the kagome layer (Fig. 3c)[53]. $Co_3Sn_2S_2$ also represents one of the first experimentally realized magnetic Weyl semimetal[54,55], a discovery that had been eagerly anticipated since 2011, when the Weyl semimetals in time reversal symmetry breaking (TRSB) systems, such as iridates, were predicted[56]. Moreover, optical and transport studies have revealed unconventional electromagnetic responses at magnetic domain walls, suggesting unique topological dynamics[57-59]. The ability to manipulate domain structures with ultrafast optical techniques, including chirality switching and domain-wall-selective transport, positions $Co_3Sn_2S_2$ as a promising candidate for topological spintronic applications[59-61].

### 1.3 $AV_3Sb_5$

Among the extensive family of kagome materials, the $AV_3Sb_5$ (*A*: Cs, K, and Rb) has garnered significant attention due to the coexistence of multiple intertwined quantum states, including CDW, pair density wave (PDW), superconductivity, nematic order, loop current, and $Z_2$ topology[14-17,62,63]. Notably, these phenomena are not merely indicative of individual quantum states; rather, they are intricately intertwined with one another due to the presence of vHS close to the $E_F$ (Fig. 3f)[7,62-64]. The $AV_3Sb_5$ series exhibits several structural characteristics. Under ambient conditions, the material adopts a crystal structure with a space group of P6/mmm, featuring a pristine two-dimensional kagome network of V atoms[65]. The kagome layer is part of a multilayered structure that alternates between Sb layers and alkali-metal layers (Fig. 2). This arrangement enriches the electronic structure by providing multiple degrees of freedom, enabling the emergence of unconventional CDW below 95 K and unconventional superconductivity below 2.5 K in $A$V$_3$Sb$_5$[14-17,62-65]. Electronic structure calculations revealed that the vHS lie close to the $E_F$ and arise primarily from V 3$d$ orbitals, which strongly hybridize with Sb 5$p$ states[66,67]. This hybridization plays a critical role in CDW gap formation and influences Fermi surface nesting. A distinctive feature of this series is the presence of multiple vHSs, which are classified into two categories: $p$-type (with a single sublattice) and $m$-type (with a mixed sublattice)[7,68]. These singularities are associated with distinct saddle points along the $\Gamma$-$M$ direction and are particularly prone to electronic instabilities. The CDW primarily shows how to reconstruct the Fermi surface and break rotational symmetry (nematicity) while preserving translational symmetry[16]. This suggests the presence of a bond-order wave rather than a traditional Peierls instability. The phenomenon is most evident in the manifestation of diverse stacking in the $c$-axis: 2×2×2 (Ref.[69-72]), 2×2×4 (Ref.[73,74]) and even coexistence of both order in CsV$_3$Sb$_5$, thereby supporting multiple motifs, including the Star of David, the tri-hexagonal (inverse Star of David), and 4Q orders (Fig. 3h)[71,75]. Importantly, the CDW gap opens near the vHSs, whereas the Dirac points remain largely unaffected[17,68]. The CDW mechanism is not purely driven by nesting, but also reflects the role of electron correlations and topological band features[14]. The CDW exhibits the following characteristics: a broken $C_2$ symmetry, TRSB, chiral charge order, and domain-boundary phase shifts[16,76-80], resulting in intriguing phenomena such as the superconducting diode effect and nonreciprocal transport (Fig. 3d)[18,81]. In addition, the $A$V$_3$Sb$_5$ series exhibits unconventional superconductivity intertwined with PDW, nematicity, and CDW orders[14,16,62]. The application of external pressure suppresses the CDW phase near 2 GPa and simultaneously enhances superconductivity[64]. The strain and doping (e.g., Nb, Ti, Sn) have been shown to induce double-dome superconductivity, which is nodeless multiband linked to the CDW state[63,82-86]. Signatures of short-range PDW order have been linked to small hole pockets induced by the CDW phase[87]. In addition to these phases, the series displays nematic order, attributed to spontaneous TRSB and loop current states, which may give rise to switchable chiral charge order without net magnetization[81,88-91]. The origin of TRSB in the CDW phase remains a subject of considerable debate[92]. These quantum phases are highly sensitive to chemical substitution (e.g., Ti, Nb, Mo, Cr, Sn) and interlayer shifts, which significantly influence the CDW coherence and the superconducting state[85,86,93-96].

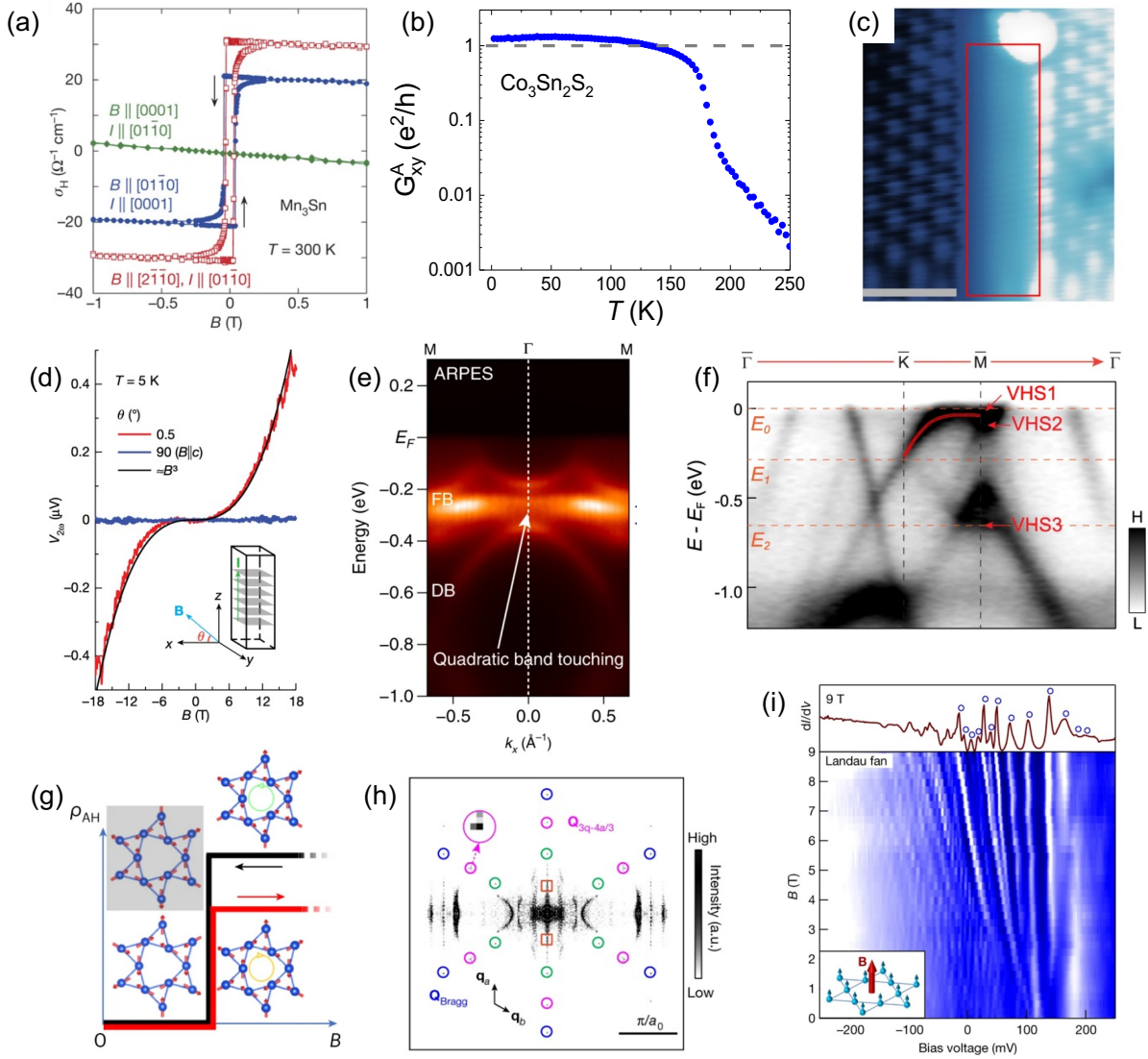

**Fig. 3 Various key phenomena in kagome quantum materials. a** Hall conductivity ($\sigma_H$) of $Mn_3Sn$ at 300 K, adapted from Ref.[20]. **b** Quantized Hall conductance of $Co_3Sn_2S_2$, adapted from Ref.[4]. **c** Topography of 1D chiral edge states at the terrace of $Co_3Sn_2S_2$, adapted from Ref.[53]. **d** Second-harmonic voltage varying with $B^3$, revealing a chiral nature of CDW in $CsV_3Sb_5$, adapted from Ref.[81]. Inset shows the angle between applied current and field. **e** ARPES showing Dirac band and FB around the $\Gamma$ point of CoSn, adapted from Ref.[45]. **f** van Hove singularities and Dirac bands (red line) of $CsV_3Sb_5$, adapted from Ref.[7].The orange dashed lines indicate energy levels. **g** Illustration of the finite-field hysteresis of the anomalous Hall resistivity size in HoAgGe due to emergent time-reversal-like degenerate states, adapted from Ref.[97]. **h** $2a_0 \times 2a_0$ charge density wave (CDW) order ($Q_{3q-2a} = 1/2Q_{Bragg}$), $4a_0$ CDW order ($Q_{1q-4a} = 1/4Q_{Bragg}$) and pair density wave (PDW) order ($Q_{3q-4/3a} = 3/4Q_{Bragg}$), measured by scanning tunnelling microscopy (STM) and scanning tunnelling spectroscopy (STS) in $CsV_3Sb_5$, adapted from Ref.[62]. **i** Landau fan diagram of the kagome lattice of $TbMn_6Sn_6$. The top panel shows the 9 T $dI/dV$ spectrum, with Landau levels marked by open circles, adapted from Ref.[98].

## 1.4 $RM_6X_6$

A comprehensive investigation of the $A$V$_3$Sb$_5$ systems has revealed the recent advent of the $RM_6X_6$ series of kagome systems, which represent a substantial subset of a vast array of compounds[99,100]. These systems offer a high degree of chemical tunability, which is enabled by a multitude of available options. These compounds are characterized by hexagonal symmetry ($P6/mmm$), wherein kagome layers composed of transition metals $M$ (e.g., Mn, V, Fe, Cr) and $R$ (rare-earths or Li) are interwoven with $X$ (Sn, Ge) layers (Fig. 2). The structure and magnetism are closely coupled, and the system is composed of quasi-2D transition metal kagome layers that are stacked along the $c$-axis. The magnetism governs the interaction between the $R$ and kagome lattice atoms[101]. In particular, the $R$ atom occupies a pivotal position within the crystal structure of $R$Mn$_6$Sn$_6$, situated at the intersection of the kagome and the conventional hexagonal lattices. This strategic location enables a direct interaction between the $R$ atom and the Mn $3d$ electrons, thereby facilitating a complex electronic coupling[102]. This phenomenon enables the realization of highly tunable magnetic engineering in the series[103-105]. In the context of $R$Mn$_6$Sn$_6$ compounds, the Mn-$R$ interaction exhibits a predominance that surpasses the Mn-Mn interaction[101]. This observation leads to the concurrent ordering of both Mn and $R$ atoms, a phenomenon that persists even at ambient temperatures. However, the transition towards nonmagnetic rare-earths (Sc, Y, Lu), the coupling between $R$-Mn vanishes, resulting in a more complex magnetic behavior that is influenced by both temperature and field. For instance, YMn$_6$Sn$_6$ manifests collinear AFM order at temperatures (T < 350 K)[106]. The Mn moments align ferromagnetically in the plane of the kagome layers, yet antiferromagnetically between adjacent layers (along the $c$-axis). However, upon substituting Sn with Ge, the prevalence of rare-earth elements diminishes, as evidenced by a decline in ferrimagnetic ordering[107-109]. In this configuration, the magnetic moments undergo a reorientation of approximately 40 to 60 degrees as they transition from one $R$ sheet to another[110]. Also, moving towards Fe-based 166 compounds show strong antiferromagnetic ordering due to strong coupling between Fe-Fe kagome layers[107].

The presence of a rich magnetism, topological band structures, and correlations provide a plethora of topological properties with respect to topological signatures[98,111-117]. Many compounds have been observed to exhibit collinear ferromagnetism at temperatures above ambient, suggesting the potential for AHE and ANE behavior[112,118]. In the case that the topological bands are close to the $E_F$, the intrinsic AHE becomes dominant, with $R$Mn$_6$Sn$_6$ compounds serving as primary systems to illustrate this phenomenon[118]. Depending on the $R$-$M$ magnetic interactions, some compounds exhibit a topological Hall effect (THE), which indirectly suggests the presence of real-space spin textures[111,116]. In the case of TbMn$_6$Sn$_6$, a ferrimagnetic ground state is observed at temperatures below 420 K, where the Tb magnetic moment is antiparallel to the Mn lattice, and the material manifests pronounced out-of-plane magnetization[98]. This indicates it as a promising material for the study of topological phenomena, including Chern gapped Dirac fermions[119]. The electronic states demonstrate clear Landau quantization in response to the application of a magnetic field, aligning with the characteristics of Chern bands (Fig. 3i)[98]. Consequently, $TbMn_6Sn_6$ functions as a prototype system to investigate the interplay between topology, magnetism, and lattice dynamics. Its sister compound, $HoMn_6Sn_6$, exhibits anomalous magnetic and transport properties, including a substantial AHE and a Chern gap[120,121].

As is the case with $A$V$_3$Sb$_5$, an unconventional CDW also manifests in $ScV_6Sn_6$ at 92 K, accompanied by anomalous Hall-like behavior, which persists up to the CDW[122-124]. The CDW is driven by the displacement of ~ 0.15 Å of Sc and Sn atoms compared to the displacement of ~ 0.005 Å of V. It is the opposite of CDW in the $A$V$_3$Sb$_5$ system, where V atoms displace the most. Thus, the CDW is more three-dimensional, involving rhombohedral distortion, and is reported to be $\sqrt{3} \times \sqrt{3} \times 3$(Ref.[125]). It is noteworthy that the absence of a CDW gap in proximity to a vHS suggests a distinct mechanism compared to the instability observed in the vicinity of vHS and is associated with the softening of a zone-boundary phonon with $q = (1/3, 1/3, 1/3)$(Ref.[126-128]). Subsequent to the application of internal chemical pressure through substitution or external pressure, CDW undergoes a gradual transition to low temperatures, ultimately leading to its complete dissolution without any indication of superconductivity[129]. Nevertheless, there are still a considerable number of unanswered questions regarding the unconventional CDW of $ScV_6Sn_6$, such as the origin of TRSB and AHE.

## 1.5 Distorted kagome

Distorted kagome compounds are materials in which the ideal, perfectly symmetric kagome lattice structure is altered, often by rotating or twisting the kagome triangles that comprise the lattice. This distortion disrupts the inversion symmetry, which can result in anomalous physical properties, including unconventional magnetic behavior and the emergence of new topological phenomena[25,130]. These distortions may arise from atomic displacements, chemical substitutions, or lattice anisotropy, resulting in modified bond lengths and angles that alleviate the geometrical frustration intrinsic to the ideal kagome lattice. A considerable number of series, predominantly $R$Ti$_3$Bi$_4$, Nb$_3$$X$$_8$ ($X$: Cl, Br, I), and $RMX$ ($R$: rare-earths, $M$: transition metals, $X$: Sn, Ge, P), have been identified as exhibiting this phenomenon[130-132]. The distortion in these series can initially be categorized based on the plane in which it occurs. For instance, the distortion manifests in the out-of-plane direction of the kagome lattice in $R$Ti$_3$Bi$_4$, while it occurs in the in-plane direction in Nb$_3$X$_8$ and $RMX$ series[130-132]. However, $RMX$ demonstrates distortion through alterations in bond angle, while Nb$_3$X$_8$ exhibits distortion by changes in bond length[130,132].

$R$Ti$_3$Bi$_4$ crystallizes in an orthorhombic structure with $Fmmm$ space group. This series exhibits weak interlayer coupling, a property that facilitates exfoliation and manifests as a spectrum of magnetism[131]. This magnetism is governed by a zigzag rare-earth lattice, exhibiting a transition from ferromagnetic SmTi$_3$Bi$_4$ to AFM Gd/TbTi$_3$Bi$_4$ and paramagnetic YbTi$_3$Bi$_4$. However, GdTi$_3$Bi$_4$ is distinctive due to its charge-spin intertwined density wave and field controlled striped plateau phases[131,133]. It is noteworthy that EuTi$_3$Bi$_4$ exhibits robust topological surface states that persist throughout its magnetic transitions[134]. Across the family, rare-earth substitution and pressure-modulated magnetic order and electronic bandwidth are responsible for the control over band topology[135]. Conversely, the plane distortion in breathing kagome lattices (Nb$_3$X$_8$) offers a promising avenue for the exploration of topological flat bands, quantum spin liquid, and potential triferroic states in exfoliable two-dimensional semiconductors[136-138]. These lattices are distinguished by their alternating triangle sizes. Nb$_3$Cl$_8$ exhibits a charge-spin entangled cluster mott state, while its Br counterpart maintains a weak dispersion and ambient stability[139-142].

The following discussion will address angular distortion in the $RMX$ series[130,143]. These compounds represent a complex magnetism with non-trivial topology and often realized in $P\bar{6}2m$ symmetry (Fig. 2). The complex magnetism exhibited by HoAgGe is distinguished by its distinctive structure, which manifests as a spin ice state and a time-reversal like symmetry breaking phenomenon (Fig. 3g)[24,25,97]. This observation is particularly noteworthy as it is observed in the electronic transport characteristics, a phenomenon that is rare. Additionally, it has been observed that distortion enhances the AHE by field-induced Weyl points in antiferromagnetic compounds[25]. Recent theoretical explorations of such systems have revealed the emergence of pure electronic chirality, chiral phonons, and phononic surface states, phenomena that require experimental verification[144-146].

## 1.6 Synthesis of kagome materials

Recent advancements in kagome quantum materials have given rise to novel methodologies for the observation and comprehension of quantum phenomena. While these materials exhibit exotic properties, such as charge order, complex magnetism, unconventional superconductivity, and topology, the manifestation of these phenomena is contingent on the quality of the bulk single crystals. For a recent example, structural distortion in FeGe has been demonstrated to be tunable by annealing, thereby enabling the modulation of both magnetic ordering and CDW[147]. Similarly, a small disorder can significantly impact the physics of the $A$V$_3$Sb$_5$ series[148]. These previous investigations demonstrate the hypothesis that a physical property is contingent on the quality of the crystal. Single crystals are composed of a continuous, unbroken three-dimensional atomic lattice with a uniform repeating structure throughout. The synthesis of such crystals is frequently laborious and protracted. Nevertheless, their significance lies in the clarity when probing intrinsic physical properties, as they are devoid of the complications introduced by grain boundaries and secondary phases. Furthermore, single crystals are indispensable for the study of anisotropic behaviors, as they allow for precise measurements along distinct crystallographic directions, thereby reflecting inherent structural anisotropies. These properties render single crystals indispensable for the precise characterization of quantum materials by utilizing several techniques. Depending on several factors, including volatility, thermodynamic stability, the extent of doping, choice of the phase, *etc*., the selection of crystal growth techniques depends. The popular and extremely useful methods are solution growth (metal-flux), vapor phase growth (chemical vapor transport (CVT)), and melt growth (Bridgman, floating zone (FZ), and Czochralski (CZ)) for growing single crystals of a variety of quantum materials (Table 1)[149,150].

The kagome materials are generally grown by the flux method due to being intermetallic in nature and containing low melting point elements such as Sn, In, Sb, and Bi, as mentioned in Table 1. $A$V$_3$Sb$_5$ system is grown by the self-flux method in which Sb acts as both reactant and flux[65]. Similarly, in $R$M$_6$Sn$_6$, Co$_3$Sn$_2$S$_2$ and $M_m$Sn$_n$ series, Sn acts as both reactant and flux[20,52,118]. Unlike a single element as a flux, the eutectic composition of a binary can also be used as a flux, such eutectic point of Ag-Ge composition is used for growing the $R$AgGe series[130]. In this typical procedure of the flux method, elements of respective materials are taken in an appropriate amount by considering the solubility phase diagram of elements in flux element (e.g., in a molar ratio of A: V: Sb = 1:3:14)[65]. After putting them in a suitable crucible (e.g., alumina), the crucible is then sealed in an evacuated quartz ampoule. The mixture is then heated to a high temperature to ensure that all elements are in a liquid state. It is then maintained at this temperature for a period of 24- 48 hours for homogeneity of the liquid. Thereafter, the liquid is gradually cooled at a rate of 2 to 5°C/h to a lower temperature, which is typically above the melting point of the flux and below the solidification temperature of the crystals. Following this cooling process, the centrifuge procedure is performed. In this procedure, nice crystals are grown whose facets are visible. However, there are several challenges that need to be considered while selecting the crystal, including physical or chemical binding of the crystals with the used flux and incorporation of the flux inside the crystal as an impurity.

The most common issue with flux growth is the intercalation of flux between the layers of a single crystal. This greatly influences the physical properties of the materials. In the case of $A$V$_3$Sb$_5$ series, it has been observed that even small variations in stoichiometry or disorder can significantly impact the already complex physics of the series[148]. Multiple factors contribute to the challenges associated with the flux growth, including but not limited to: residue flux, a notably slow cooling rate, the difficulty in regulating crystal size, the complexity of achieving high purity, crucible contamination, and scalability issues. Furthermore, the process of centrifugation at elevated temperatures poses significant safety hazards.

**Table 1 | Property, growth, and challenges of kagome materials**

| Compound | Growth method | Key property | Key challenge | Reference |
|---|---|---|---|---|
| $A$V$_3$Sb$_5$ (A: Cs, Rb, K) | Flux (Sb self-flux) | CDW, PDW, superconductivity, nematicity, non-reciprocal transport | Flux intercalation, stoichiometry control, volatile elements | 16,62,65,81,151,152 |
| $R$M$_6$X$_6$ (R: rare-earths or Li; $M$: Mn, Fe, V and $X$: Sn, Ge) | Flux (Sn or Ge self-flux) | CDW (ScV$_6$Sn$_6$), quantum Chern phase (TbMn$_6$Sn$_6$), AHE and THE | Flux intercalation, phase separation, magnetic ordering control | 98,101,105,111,122 |
| Co$_3$Sn$_2$S$_2$ Co$_3$In$_2$S$_2$ | Flux (Sn) and CVT (I$_2$-transport agent) | Magnetic Weyl semimetal, large intrinsic AHE | Flux contamination, impurity segregation | 52,54 |
| $M_m$X$_n$ ($M$: Co, Mn, Fe; $X$: Sn, Ge & $m, n$ = 1-3) | Flux (Sn/ Ge), Brigeman (Mn$_3$Sn, Mn$_3$Ge) | Non-collinear AFM, room-temperature AHE, chiral magnetic structure | Flux residue, phase purity | 20-22,41,153,154 |
| $R$Ti$_3$Bi$_4$ (R: rare-earths) | Flux (Bi-flux) | CDW, SDW, easily exfoliable along c-axis | Flux residue between layers | 131,133 |
| Nb$_3$X$_8$ ($X$: Cl, Br, I) | CVT ($X$-transport agent) | Topological surface states, Mott insulator, triferroic | Layer stacking faults, growth window sensitivity | 136,142,155,156 |
| $RMX$ (R: rare-earths, $M$: transition metals, $X$: Sn, Ge, P) | Flux (mixed flux for e.g., Ag-Ge) | Spin ice state (HoAgGe), complex magnetism | Hard removal of the flux | 24,97,130 |

One of the most versatile methods for growing crystals of intermetallic compounds is the Bridgman technique, which can be used even if the melting point of the elements is relatively high. This method involves a vertical crucible containing the melt, which moves slowly relative to a temperature gradient. When the crucible leaves the hot zone, solidification takes place at the bottom. It is essential to nucleate a very small part of the sample to reduce the number of grains. A pointed-tip-shaped crucible is used to achieve this. Bridgman-grown crystals are rod-shaped ingots. However, it is noteworthy that these ingots frequently contain a mixture of different grains. It is strongly advised that Ta tubes be utilized in place of quartz tubes when employing high-temperature growth techniques that exceed 1150°C. For example, in the case of $Mn_3Sn$ or $Mn_3Ge$, high-purity elements with stochiometric composition are loaded in a pointed-tip $Al_2O_3$ crucible, which is sealed in Ta crucible and then heated to at least 100°C above the melting point of the respective compound to ensure complete loaded quantity is in a liquid state[21,153]. This method is optimum for congruent melting phases. However, off-stoichiometry is also used to obtain an equilibrium phase in particular cases, and an extra amount of component is segregated on the side of the rod.

The aforementioned methods are the most viable approach for growing single crystals. However, it should be noted that these methods do not ensure the desired quality. Consequently, there are numerous strategies that can be employed to overcome these challenges. For instance, the CVT method for $Co_3Sn_2S_2$ and $Co_3In_2S_2$ yields crystals of higher purity. This is evidenced by the observation of quantum oscillations and ultra-high residual resistivity ratio ($RRR$) value (~1600) of magnetic semimetal compounds[52,157]. Conversely, alternative methods, such as FZ, Bridgman and Czochralski methods, can be employed to achieve flux-free growth, enhance the dimensions of single crystals, and reduce the growth time. However, these methods also give rise to new challenges and should not be applied universally. It has been observed in numerous materials that two phases are thermodynamically stable in close proximity to each other, and it is challenging to separate one from another during the process of growth. For example, the growth conditions for kagome superconductor either $LaRu_3Si_2$ or $LaIr_3Si_2$ and either phase $LaRu_2Si_2$ or $LaIr_2Si_2$ are found to be quite close. Similarly, FeGe does not exist in molten form and it cannot be grown by the melt growth method. Therefore, the CVT method is the only viable option[46,158].

## 2. Chiral materials

Chiral compounds are materials characterized by a lack of mirror symmetry, meaning they cannot be superimposed onto their mirror image, much like left and right hands. In the context of condensed matter physics and materials science, chirality can also manifest in crystal structures, magnetic textures, and electronic properties[5,6,27,33,159]. Chiral crystals, for instance, have been shown to host a variety of exotic phenomena, including natural optical activity, non-reciprocal transport effects, spin-selective transport, and topologically protected states such as chiral spin textures or skyrmions. The interplay between structural chirality and electronic or magnetic degrees of freedom in these materials has become a subject of considerable interest for quantum information processing and novel device applications. The search for materials that host exotic topological states has seen rapid expansion in recent years. Topological insulators, Dirac and Weyl semimetals, and skyrmion-hosting magnets are prominent examples of such materials.

### 2.1 MX

Binary B20 *MX* chiral crystals (*M* = Mn, Fe, Co, Pd, Pt, Rh; and *X* = Al, Si, Sn, Ga, Ge) have garnered significant interest due to easy to synthesis. The crystal structure belongs to the space group $P2_13$, which supports both right- and left-handed crystal forms (Fig. 4a)[5,27,29,159,160]. The handedness of the structure is determined by the twist direction of the *M* and *X* atoms in the crystal, with overall crystal chirality defined by the handedness of the *X* helices. Recent studies have thoroughly investigated PdGa, PtGa, CoSi, RhSi, and PtAl for Weyl fermions, multifold fermions, and nontrivial surface states. These properties render the B20 materials optimal candidates for the study of unconventional topological quasiparticles that are beyond the conventional Dirac and Weyl fermion paradigms. It is worth mentioning that a new class of fermions, the Kramers-Weyl fermions, appear at high-symmetry points in the Brillouin zone and are protected by symmetry. They are connected by arc-like Fermi states that span across a large part of reciprocal space (Fig. 4b)[161]. These fermions are stabilized by structural chirality and SOC, without the need for band inversion. Kramers-Weyl fermions can have three-fold, four-fold, or six-fold crossings, which are associated with higher Chern numbers and higher-order fermions (Fig. 4c)[28]. Hundreds of materials in 65 chiral space groups are potential hosts for Kramers-Weyl fermions, with some already experimentally observed- PdGa and CoSi being a notable example[5,31,161]. A monopole-like orbital-momentum locking spins texture on the three-dimensional Fermi surfaces has been observed, leading to a large orbital Hall effect (OHE) and a giant orbital magnetoelectric effect. Importantly, while the OHE remains consistent across different enantiomers, the orbital magnetoelectric effect is chirality-dependent and surpasses its spin counterpart in magnitude. The study highlighted the crucial role of orbital texture in understanding these effects, paving the way for applications in orbitronics and spintronics[159]. Orbital angular momentum monopoles in PtGa and PdGa reveal a polar texture that rotates around the monopole with changing photon energy, arising from the magnetic orbital texture and atomic contributions. The polarity of monopoles can be controlled by the crystal's handedness, as demonstrated in the enantiomers of PdGa, where monopoles and antimonopoles are imaged[29]. Weyl spin-momentum locking is observed in all types of Weyl semimetals but it has been resolved clearly in PtGa due to strong SOC induced splitting of bands. The results demonstrate that the electron spin in the Fermi arc surface states is orthogonal to the Fermi surface contour near the bulk multifold fermion's projection at the $\Gamma$ point[162]. Similar results of the bulk spin texture locking at the $R$ point are also visible[161,163]. Recent studies have demonstrated that the chirality of B20 crystals, such as CoSi induces distinct orbital angular momentum (OAM) textures in their bulk electronic structures. Using circular dichroism in soft X-ray ARPES, researchers observed that these OAM textures are directly linked to the crystal's handedness[28,159,164]. ARPES study also identifies two unconventional chiral fermions: spin-1 and charge-2 fermions in CoSi and RhSn crystals[159]. These fermions, enforced by crystal symmetries, are connected by long Fermi arcs across the surface Brillouin zone, differing from typical Weyl semimetals[161]. The findings confirm the existence of these unconventional chiral fermions and open new opportunities to explore their unique physical properties. In addition, optical conductivity studies have also been utilized to identify multifold fermions[163,165]. The further study of chiral materials focuses on achieving precise control of electron spin at a surface that exhibits intrinsic chirality. This control is critical for enhancing the efficiency of the oxygen reduction reaction (ORR), exploring how manipulating electron spins on such surfaces can significantly improve catalytic activity (Fig. 4d)[166,167].

A famous magnetic counterpart of the B20 structure compound is MnSi, which exhibits a helical magnetic state. Such a state is stabilized by ferromagnetic exchange and weaker Dzyaloshinskii-Moriya interaction (DMI), which is allowed by the non-centrosymmetric crystal structure[33]. The magnetic helix propagates along the [111] directions of the unit cell, but its periodicity is much longer than the atomic lattice due to weak coupling between the magnetic and atomic structures. A distinct phase, called the "*A phase*", is observed at low fields just below the $T_C$, where topologically-protected magnetic vortices known as skyrmions are formed (Fig. 4f)[33,168]. This study has provided a particularly illustrative example of the topological arrangement of spins, thereby contributing to the scientific understanding of the subject. The existence of skyrmions is further confirmed by the signature of THE observed in magneto-electrical transport measurements[169]. Another feature of MnSi is that it exhibits chiral spin fluctuations through the electrical magnetochiral effect[170]. Significant magnetochiral signals are observed in specific regions of temperature, magnetic field, and pressure: near the helical ordering temperature in the paramagnetic phase and in the partially ordered topological spin state at low temperatures and high pressures. These findings highlight the potential for new electromagnetic functionalities in chiral magnets.

In addition to the anticipated existence of multifold fermions, Weyl spin-momentum locking and skyrmions in B20 chiral compounds, several other properties are expected to result from the presence of structural chirality. A second harmonic generation (SHG) over a wide frequency range is found in RhSi, contributing from multifold fermions[165]. It demonstrates that transitions between linearly dispersing bands, especially near topological band degeneracies, actually suppress SHG rather than enhance it. In contrast, two-photon transitions between other dispersive bands, particularly those with flatter dispersions, dominate SHG responses. Furthermore, a many-body effect that favors two-photon transitions contributes to this suppression. Intrinsic nonlinear planar Hall effect (INPHE) emerges from the interplay between structural chirality and topological nodal fermions, resulting in a Hall response that is sensitive to the material's handedness. This phenomenon is attributed to diverging orbital magnetic moments with hedgehog-like textures around the nodes, resulting in a conductivity that is proportional to the topological charge. It is noteworthy that materials such as CoSi and PtAl have been proposed to exhibit a giant INPHE conductivity (1 to 10 A $V^{-2}$ $T^{-1}$), rendering them promising candidates for applications in enantiomer recognition and nonlinear transport devices[171]. Furthermore, the quantum oscillations in the chiral materials reveal a quasi-symmetry protection, which is broken by higher-order perturbations. This gives rise to finite but parametrically negligible energy gaps at specific points in momentum space[172]. Conversely, the manifestation of crystal symmetries gives rise to two-dimensional band degeneracies (nodal planes) across the BZ in ferromagnetic MnSi. This phenomenon is a direct consequence of the material's nonsymmorphic symmetries, particularly screw rotations combined with time-reversal symmetry, which are inherent to its chiral crystal structure[173].

## 2.2 $M_{1/3}M'X2$

Transition metal dichalcogenides (TMDs) are a class of layered materials with the general formula $MX_2$, where $M$ is a transition metal (e.g., Mo, W, Ti) and $X$ is a chalcogen (S, Se, or Te). The structure of these materials consists of covalently bonded $X$–$M$–$X$ layers, held together by weak van der Waals forces, enabling facile exfoliation down to monolayers[174]. This series of compounds gives rise to a diverse array of electronic behaviors, encompassing the phases of semiconductors, metals, and superconductors, and have already been proven to show quantum properties, such as Weyl semimetals[175], topological edge states[176], quantum spin Hall effects[177,178], and Ising superconductivity under certain conditions[179]. The tunable properties of TMDs, which can be modulated by layer number, strain, gating, or doping, in conjunction with strong SOC and their two-dimensional (2D) nature, position them as a versatile platform for exploring novel quantum phenomena and developing next-generation nanoscale devices. The intercalated TMDs represented by the formula $M_{1/3}M'X_2$, (where $M'$ = Nb, Ta) is facilitated the formation of complex magnetic textures, depending on the type of transition metal used as the intercalants[180-186]. On this end, compound $Cr_{1/3}NbS_2$ displays a helical magnetic ground state below its $T_C \sim 130$ K. It adopts a hexagonal structure with either $P6_322$ or $P6_3$ symmetry, depending on Cr ion ordering. In zero a magnetic field, $Cr_{1/3}NbS_2$ has a chiral helical magnetic ground state, with spins confined to the $ab$-plane and helices propagating along the $c$-axis. This helical arrangement results from a combination of magnetic anisotropy from crystal fields, intralayer ferromagnetic exchange and DMI. Under applied magnetic fields, the system transitions from helical to conical spin states (along the $c$-axis), eventually becoming fully ferromagnetic at high fields [183]. Most recently, another nontrivial, one-dimensional spin texture known as a chiral soliton lattice (CSL) has been observed to form in this material within the $ab$-plane, where ferromagnetic regions are separated by kinks (solitons). The size of these regions increases with field strength until a critical field induces a ferromagnetic state. It seems that such CSL produces a nonreciprocal flow of conduction electrons, giving rise to the electrical magnetochiral (EMC) effect, which exists over a wide range of magnetic fields and temperatures. In the presence of a chiral conical phase, a significant enhancement of the second harmonic resistance is observed and the value of the resistance is three orders of magnitude greater than the contribution from crystal chirality[182]. Isostructural $Cr_{1/3}TaS_2$, is also a layered monoaxial chiral helimagnet ($T_C \sim 110$ K) with a tunable heterochiral state. It shows a macroscopic spiral-like magnetic texture within each domain, formed by quasiperiodic Néel domain walls. Interestingly, the spirality of these textures can be either sign and is independent of the structural chirality. Under weak in-plane magnetic fields, these spirals transform into concentric ring domain patterns[180,184]. Furthermore, Co and Fe intercalations lead to the usual collinear AFM and FM phases. Particularly, $Co_{1/3}MS_2$ ($M$ = Nb, Ta) exhibits spontaneous Hall effect despite exhibiting negligible net magnetization (Fig. 4h), yet the microscopic origin remains ambiguous[185,186]. Conversely, $Fe_{1/3}TaS_2$ constitutes a simple metallic ferromagnet, wherein chirality is superimposed upon a complex free carrier response from both Ta and Fe bands. The presence of a honeycomb charge density pattern in the Fe layer and a hole-to-electron pocket crossover at the K-point has been observed[181].

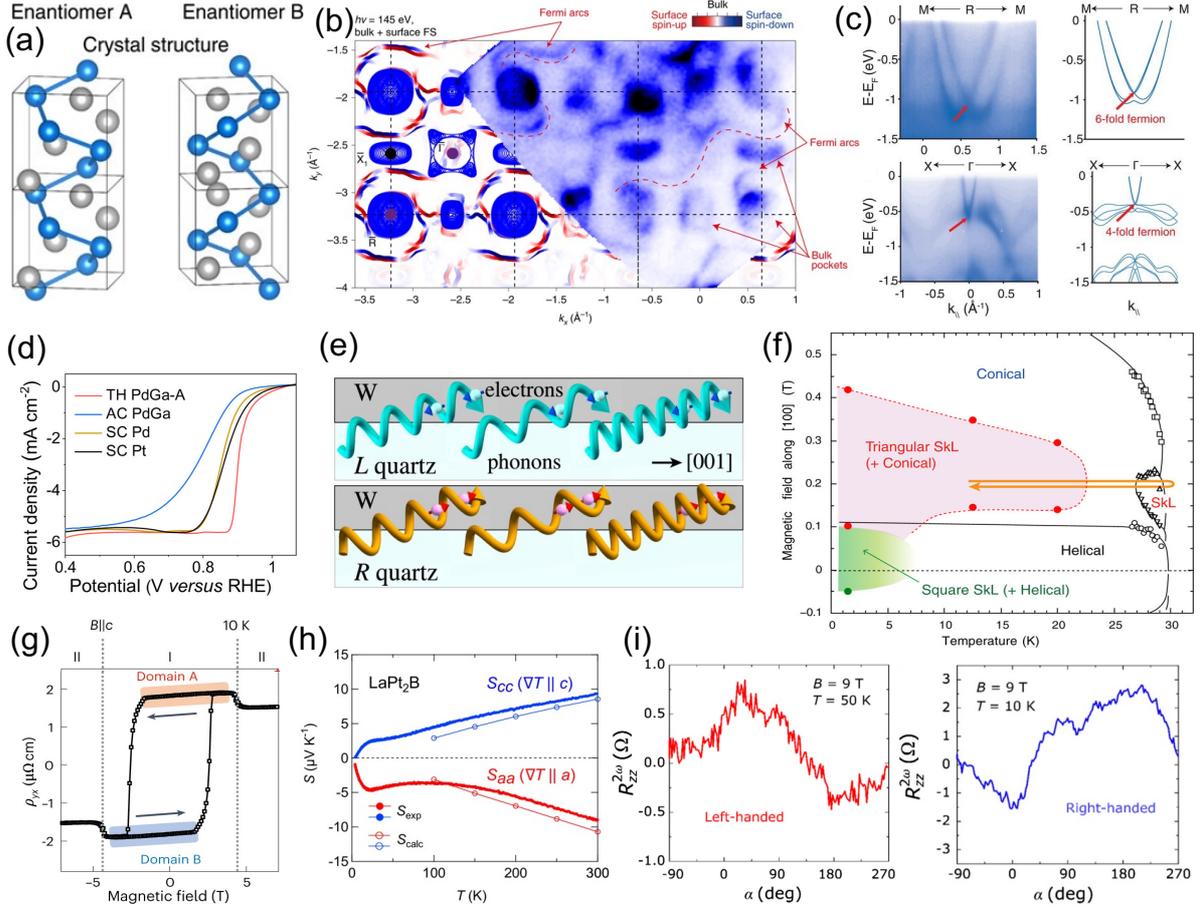

**Fig. 4 Various key phenomena in chiral quantum materials. a** chiral arrangement of $X$ atoms in both handedness. $M$ and $X$ atoms are denoted by blue and grey spheres, respectively. **b** ARPES spectra and calculated bands of PtAl showing a long Fermi arc extending across the whole Brillouin zone, adapted from Ref.[161]. **c** ARPES spectra showing multi-fold fermions in PdGa, adapted from Ref.[28]. **d** Demonstration of superior ORR activity of homochiral PdGa, exhibiting the highest onset potential, adapted from Ref.[167]. **e** A schematic propagation of phonons interacting with electrons in the α-quartz crystal, adapted from Ref.[187]. **f** Magnetic phase diagram of triangular & square skyrmion lattices and conical & helical magnetic states of MnSi, adapted from Ref.[168]. **g** Hall resistivity for $B \| c$ at 10 K for $Co_{1/3}TaS_2$, adapted from Ref.[186]. **h** Large Seebeck coefficient measured along $a$- and $c$-axes for $LaPt_2B$, adapted from Ref.[188]. **i** Electrical magnetochiral effect in elemental Te, adapted from Ref.[32].

## 2.3 RPt₂B

$RPt_2B$ belongs to a chiral kagome series of compounds, which crystallizes in a hexagonal structure in either $P6_222$ or $P6_422$ chiral space group. To date, the series of $RPt_2B$ compounds has been reported for Y, La, Ce, Pr, Nd, Tm, Gd, Yb, and Lu. *R* ions typically arrange themselves in a helical configuration when viewed along the *c*-axis. The *R*-Pt and B layers are stacked alternately along the *c*-axis, serving as an illustration of the emergence of novel spin textures within the context of a chiral crystal structure[188-190]. In another aspect, the Pt atoms form the kagome network, while the *R* and B atoms occupy interstitial sites within the kagome network. Unlike the B20 cubic system, hexagonal chiral crystals possess a single principal helical axis and are thus classified as monoaxial chiral, while *R* atoms form a 3D kagome structure. As previously mentioned, the chirality influences the manner in which a material interacts with spin and symmetry-breaking effects, while the kagome lattice governs geometrical frustration and band topology. The combination of chirality with the 3D kagome lattice results in the surprising observation of one-dimensional behavior on the surface. This behavior manifests as topological surface Fermi arc states that connect Weyl fermions, exhibiting dispersive characteristics in one momentum direction and a flat behavior in the other. These 1D Fermi arcs open up unique possibilities for generating unconventional non-local transport phenomena at the interfaces of domains with different handedness, and the associated enhanced conductance as the separation of the leads on the surface is increased[191]. In recent developments, the symmetry-breaking effects induced by the crystal structure have given rise to intriguing spin-polarized states and topological invariants, which manifest in the form of Weyl points or Dirac nodes, depending on the specific material composition and the rare-earth ion[192]. In *R*Pt$_2$B chiral crystals, the combination of crystal symmetry, strong SOC (due to the Pt atoms), and electronic interactions allows the formation of Weyl points or Dirac nodes. The symmetry-breaking effects induced by the chirality of the lattice enhance the formation of these topologically protected points, resulting in exotic surface states, such as Fermi arcs in the case of Weyl semimetals, which are robust against impurities and defects. This gives rise to interesting consequences for the material's electronic and transport properties, including the manifestation of highly anisotropic conductivity and the emergence of topologically protected edge states. In addition, the interplay between SOC, chirality, and the rare-earth magnetic moments leads to intriguing magnetic properties. These materials are expected to exhibit nontrivial magnetic textures, which may be tunable via external fields or pressure. In GdPt$_2$B, magnetization measurements and Curie-Weiss analyses indicate dominant ferromagnetic interactions, with a clear transition observed at approximately 87 K. This transition is evident in temperature-dependent electrical resistivity, magnetic susceptibility, and specific heat measurements. The compound's magnetic phase diagram reveals regions of field-polarized ferromagnetism and a magnetically ordered phase, akin to behaviors observed in known chiral helimagnets. In NdPt$_2$B, analogous structural chirality is evident, resulting in intricate magnetic ordering. The interplay between the crystal's chiral symmetry and the magnetic moments of Nd ions results in unique magnetotransport signatures, including the observation of Weyl points in the electronic structure. Both GdPt$_2$B and NdPt$_2$B exhibit significant AHE, demonstrating an anomalous Hall conductivity of up to 60 $\Omega^{-1}\cdot$cm$^{-1}$ and an anomalous Hall angle as large as 23%. This effect is attributed to the presence of Weyl points, which are induced by magnetic fields. These Weyl points result in the splitting of electronic bands near the $E_\mathrm{F}$[189,190]. On the other hand, the non-magnetic counterpart LaPt$_2$B shows axis-dependent conduction polarity, a phenomenon referred to as goniopolar materials (Fig. 4g). Its mixed-dimensional Fermi surfaces, combining quasi-1D hole sheets and quasi-2D electron sheets, give rise to axis-dependent thermopower. LaPt$_2$B shows an exceptionally large transverse Peltier conductivity (up to 130 A K$^{-1}$ m$^{-1}$), outperforming topological magnets that rely on the anomalous Nernst effect. These findings highlight mixed-dimensionality as a crucial feature for efficient transverse thermoelectric devices[188].

## 2.4 Beyond the above series
### NbSi$_2$ and NbGe$_2$

NbSi$_2$ and NbGe$_2$ are a pair of chiral intermetallic compounds with space group either $P6_222$ or $P6_422$, in which Nb atoms form a helical chain, enabling unusual topological and spin textures in the electronic structure. Among these two, NbSi$_2$ exhibits unexpected spiral-shaped constant energy contours at specific binding energies, with rotation directions dependent on the crystal's chirality[193]. This phenomenon is attributed to the interplay between the bulk chiral atomic arrangement and surface states penetrating deep into the bulk. A high-resolution ARPES study reveals the presence of exotic chiral surface states on the (100) surface of NbGe$_2$ and vHSs, thereby supporting the existence of Weyl fermions and Fermi surface instability in the electronic structure. This is attributed to the crystal's inherent chirality. Despite the isostructural nature of NbGe$_2$ and NbSi$_2$, significant disparities were identified in their electrical and thermal properties. These discrepancies are ascribed to a van Hove-type singularity in the density of states, electron count, and SOC effects. As an illustration, the de Haas–van Alphen oscillation has been demonstrated to reveal the splitting of both electron and hole Fermi surfaces in both compounds, resulting from the antisymmetric SOC interaction in slightly different energy values[194]. However, these oscillations reveal isotropic enhancement of effective mass in NbGe$_2$, attributed to strong electron-phonon melting interaction[195]. Furthermore, a key difference is that NbGe$_2$ exhibits superconductivity below 2 K, but NbSi$_2$ does not. A crossover from Type-I to Type-II superconductivity highlights the enhanced surface superconducting critical field. This phenomenon may be attributed to the presence of a Fermi arc, which connects Kramer-Weyl nodes close to the Fermi energy[194]. NbGe$_2$ exhibits significant nonreciprocal transport properties in the superconducting state and the nonreciprocal resistance-to-normal resistance ratio is found to be $6 \times 10^5$ T$^{-1}$A$^{-1}$. This effect is highly tunable, influenced by factors such as current, temperature, and crystallographic orientation, thereby highlighting the complex vortex dynamics inherent to this material.

**α-quartz and tellurium**

α-quartz (SiO$_2$) is a classic example of a chiral crystal structure. It exists in two enantiomorphic forms, belonging to chiral space groups *P*3$_1$21 and *P*3$_2$21. The chirality of the compound is derived from its trigonal crystal structure, wherein SiO$_4$ tetrahedra are arranged in a helical pattern that lacks mirror symmetry[196]. This unique structure gives α-quartz distinct properties and plays a role in prebiotic chemistry, particularly in the context of biological homochirality. The evidence suggests that the respective handedness of α-quartz may have facilitated the preferential adsorption of enantiomers of amino acids, such as alanine. This observation indicates the potential involvement of these enantiomers in the process of chiral selection during the early Earth environment[197]. In addition to its well-known application in quartz oscillators for electronic devices, the piezoelectric properties of chiral quartz are significant for exploring ore deposits and monitoring electrical signals during seismic activities. The α -quartz is optically active in rotating polarized light due to the arrangement of its crystal lattice rather than molecular asymmetry. Very recently, it was observed that applying thermal gradients to right- and left-handed single crystals of α-quartz generates voltage signals in adjacent metal electrodes made of tungsten and platinum (Fig. 4e). This phenomenon is attributed to spin currents produced via the inverse spin Hall effect. The polarity of these voltage signals is contingent on the chirality of the crystal, indicating that thermally excited phonons carry angular momentum that is influenced by chirality, which is known as chiral phonons. These findings provide direct evidence of chirality-induced selectivity in phonon angular momentum and suggest a phonon-mediated mechanism that contributes to chirality-induced spin selectivity, a phenomenon previously observed in chiral conductors and molecules[198].

Elemental Te is a chiral semiconductor characterized by helical chains arranged in a triangular lattice, interacting via van der Waals forces. This material demonstrates chirality-dependent and highly efficient conversion of charge to spin. In addition, Te exhibits a quasi-persistent radial (hedgehog-like) spin texture. The presence of inward or outward radial spin textures around the valence band maxima along the spiral chains of Te atoms is attributed to the strong SOC interaction that occurs in the absence of inversion symmetry, consistent with the crystal structure's chirality[199]. This property is likely linked with the significantly larger unidirectional magnetoresistance (MR) observed in Te compared to other materials. Another consequence of the broken spatial inversion symmetry in chiral Te is the observation of nonlinear electrical conductivity measured in devices made from single-crystalline Te flakes. The nonreciprocal electrical transport exhibits a strong second-harmonic voltage output, which scales quadratically with the input current[32]. The observed output exhibits opposite signs in left- and right-handed 2D Te due to the distinct spin polarizations that are guaranteed by the chiral symmetry (Fig. 4i). Additionally, the emergence of Weyl properties in semiconducting Te has also been investigated, with signature of negative longitudinal MR and the planar Hall effect indicating chiral anomaly effects. Logarithmically periodic quantum oscillations further reveal discrete scale invariance related to Weyl fermions[200].

**2.5 Synthesis of chiral compound**

Chiral synthesis is a specialized area of chemistry that is renowned for its emphasis on pharmaceutical applications. This distinction is significant because one form is active and effective, while the other is inactive or harmful. To deal with the physics and material science of chiral materials, crystals of both handedness are required, which is challenging. The synthesis of chiral crystals, also known as enantioselective or asymmetric synthesis, involves techniques that favor the formation of a specific enantiomer of a compound. This is achieved by using several options, including chiral starting materials, catalysts, or ligands that induce the selectivity. One key issue is preventing inorganic substances from developing racemic-twinned crystalline grains, which can exhibit both left and right-handedness during growth. A successful method for obtaining homochiral crystals with a defined handedness involves spontaneous crystallization in a stirred solution. This technique has proven effective for synthesizing water-soluble compounds. For instance, $NaClO_3$ is an achiral molecule, but it can be crystallized into chiral crystals. When crystallized from an unstirred aqueous solution, the process typically yields an approximately equal mixture. In contrast, when the solution is stirred during the process of crystallization, nearly all of the crystals adopt the same chirality, i.e. homochiral[201]. Such study highlights that enantiopure crystals are crucial for producing unique physical properties linked to the crystals' handedness. There are several established growth methods, but the most commonly used methods are mentioned below.

**Floating zone and Czochralski methods**

These methods are well known for the growth of intermetallic compounds and are applicable for growing the *MX* series of compounds of different chirality. The availability of large, high-quality single crystals is essential for various studies in materials science, which can be achieved through these methods. The pivotal option is the implementation of the seed, which tunes the growth in a particular handedness and differs from the other conventional crystal growth techniques. For example, a right-handed CoSi crystal can be produced using a right-handed FeSi seed crystal[202]. Therefore, by using an appropriate seed crystal of desired handedness, one can obtain a single crystal with that specific handedness. The key question is how to get seeds. In order to execute this procedure, it is first necessary to grow the crystal without utilizing a seed. This grown piece of crystal can subsequently be used as a seed.

The CZ technique is particularly effective for producing large quantities of high-quality single crystals of materials that melt congruently, utilizing a tri-arc, tetra-arc or induction furnace. This method has been proved advantageous for the control of chirality during the growth of B20 chiral crystals. As the developing single crystal adopts the crystallographic orientation of its submerged seed crystal within the melt. It has been previously established that $Fe_{1-x}Co_xSi$, with x = 0.08 and 0.25, crystallizes into opposing enantiomorphs. Utilizing these compositions as seeds, the growth of MnSi single crystals can be achieved to manifest left- and right-handed enantiomorphs through the CZ process[203]. Single crystals of $RPt_2B$ are prepared using the CZ method. The single crystal was pulled from the melting ingot using a tungsten seed at a speed of 12 mm h$^{-1}$ in a tetra-arc furnace under an argon atmosphere[188,189]. Homochiral crystals of transition-metal disilicides can also be produced at relatively high temperatures (around 2,000 °C) using small crystals of specific chirality as seeds in the FZ and CZ methods[193]. However, the CZ method occasionally encounters difficulties in producing homochiral single crystals from the seed ingot due to growth conditions. The high temperatures required in the furnace may exceed the melting points of the compounds under consideration. In such extreme conditions, the chirality of the seed crystal may be compromised when the homochiral seed crystal is introduced into the liquid melt at the onset of crystal growth. In contrast, the FZ method facilitates precise temperature regulation, thereby ensuring the stability of the molten zone. The apparatus functions by means of either a halogen lamp or a high-power diode laser, both of which can be readily adjusted to varying levels of power. This stability is likely to help preserve the crystallographic characteristics of the seed crystal within the connection region. Consequently, the FZ technique provides a high degree of flexibility with respect to material combinations, chirality, and the size of the produced crystals in a crucible-free environment. To prepare the feed and seed rods used for the FZ method, the polycrystalline samples of B20 are synthesized by a standard levitation rod casting furnace. To grow single crystals, travelling speeds of the feed and seed rods have to be set the same and it is necessary to rotate the feed and seed rods in the opposite direction for homogeneous liquid. The *MX* series, including PdGa, PtGa, CoSi, and FeSi, exhibit a comparable phase diagram around a 1:1 composition and they are very easy to grow several cm length crystals using FZ method.

**Bridgman method**

It has been shown that homochiral PdGa crystals can be effectively grown utilizing a seed-controlled Bridgeman method. Initially, a polycrystalline ingot is prepared via the standard arc melting method, combining high-purity Pd and Ga metals. Then the crushed powder is placed in a bottom-cone alumina crucible and sealed in a quartz tube. To manipulate the structural chirality, the previously grown $Fe_{1-x}Co_xSi$ ($x$ = 0.08 and 0.25) is used as a seed for growing PdGa crystals. The quartz ampoule is heated to 1100 °C for 12 hours, then cooled to 900 °C with rate of 1.5 °C/h, followed by further cooling to 800 °C at 50 °C/h, and finally annealed at 500 °C for 120 hours[163]. Using seeds of different compounds invite impurity and thus it is always preferable to use seed of the same compound. It is imperative to note that the melting point of PdGa is comparatively low, and its growth can be readily achieved through a straightforward heating and cooling procedure. In such a case, it is possible to grow multiple ingots concurrently, with each batch exhibiting a distinct handedness.

**Chemical vapor transport (CVT)**

CVT method is well known for growing TMDs and is also useful for intercalated chiral TMDs, i.e. $M_{1/3}M'X_2$. This method is utilized under isothermal conditions, where the driving force is not a temperature gradient, but rather the chemical potential gradient between the gas phase and the stoichiometric composition. Halogen elements are used as a transport agent and iodine is commonly used. At first, polycrystalline powder samples of constituent elements are mixed in a stochiometric 1:3:6 molar ratio and vacuum-sealed in a fused silica ampule. The ampule is first heated multiple times between 700-900 °C with intermediate grinding for several days. Such multiple grinding and heating help to improve the homogeneity. The single crystal of $M_{1/3}M'X_2$ is grown through CVT using the pre-reacted powder. The evacuated quartz tube with $I_2$ (5 mg $I_2$ /cc) is placed in a two-zone furnace with a temperature gradient from 900°C to 850°C for a week[180]. Notably, $M_{1/3}M'X_2$ single crystals can also be prepared via single step CVT technique by taking a mixture of powders of constituent elements. This procedure yields a poor quality[204,205]. However, the previously mentioned two step CVT process generally enhances the quality and dimensions of the crystals. $NbGe_2$ crystals can also be grown using the CVT technique with $I_2$ as the transport agent. Initially, polycrystalline samples are made by heating mixed Nb and Ge powders at 900 °C for 3 days. In the next step, the preheated powder is sealed in vacuum silica tubes with iodine, and the crystals are grown over a month at 900 °C with a temperature gradient of under 10 °C. Single crystals with a typical shape of a hexagonal pyramid were obtained after ultrasonic cleaning in ethanol[194]. High-quality Te single crystals are also produced using the vapor transport technique. High-purity Te powder is placed in a quartz tube with a small amount of C powder to eliminate trace oxygen. The tube is then heated to 1000 °C and held at that temperature for 1 hour. It is then cooled at 20 °C per hour to 400 °C in the hot zone and 300 °C in the cold zone. After maintaining these temperatures for 2 weeks, the tube is slowly cooled to room temperature[32,200].

**Table 2** | Property, growth, and challenges of chiral materials

| Compound | Growth method | Key property | Key challenge | Reference |
|---|---|---|---|---|
| MnSi, NbSi$_2$, Fe$_{1-x}$Co$_x$Si | CZ, FZ | THE, skyrmion lattice, chiral magnetism, chiral multifold fermions, giant Fermi arcs, topological Kondo insulator | High temperature can compromise seed chirality in CZ; maintaining homochirality is difficult. | 170,171,193,203 |
| RPt$_2$B | CZ | Chiral helimagnetism, goniopolarity, non-reciprocal transport | Requires precise temperature control, lack of control of handedness | 188-190,192 |
| PdGa, PtGa, CoSi, FeSi | FZ, Bridgman | Multifold fermions, giant Fermi arc, Kramers-Weyl fermions | Requires precise temperature control and seed preparation | 27,28,161,166 |
| intercalated chiral TMDs | CVT | chiral helimagnetism, chiral soliton lattice, EMC effect | Lack of control of handedness, smaller crystals | 180,181,183-186 |
| NbGe$_2$ | CVT | Topological superconductivity | Long growth periods, lack of control of handedness, and smaller crystals | 194 |

The CVT method offers several advantages, particularly the ability to grow high-purity, well-ordered single crystals, which is crucial for evaluating intrinsic material properties. Its flexibility allows it to be used across a wide range of TMD compositions and intercalation types. However, it also has certain limitations. A major challenge lies in controlling chirality, as CVT typically results in random chiral domains with no inherent mechanism for selecting a specific handedness, and each piece of crystals needs to be analyzed before the further use[163]. The process is also slow and sensitive, requiring precise temperature and transport agent conditions, which can lead to inconsistent results and long synthesis times. Besides, the growth of large crystals is limited, and reproducibility across batches can also be poor. Structural defects such as dislocations, while sometimes useful for inducing chirality, may degrade material performance. To control the chirality, use of a chiral agent may be helpful. For instance, NbS$_2$ and TaS$_2$ intercalated with chiral amines have shown chiral CDW phases. The growth of 1T-TaS$_2$ under the influence of chiral iodine derivatives has led to non-centrosymmetric domain formation. Additionally, the intercalation of chiral molecules into TaS$_2$ has resulted in chiral molecular intercalation superlattices[206,207].

## 3. Square-net materials

Square lattice-based materials have emerged as a fertile ground for realizing a variety of topological states, due to their symmetry rich crystallographic motifs and their compatibility with Dirac and nodal-line electronic structures[11,34-36]. In particular, materials with square-nets of *p*-block elements such as Bi, Sb, or Si exhibit significant dispersion near the $E_F$, allowing them to host Dirac fermions and nodal loops under favorable electronic configurations[11]. Furthermore, a square lattice containing magnetic ions shows interesting magnetic properties and thereby leading to Weyl nodal physics[208-210]. As shown in Fig. 5a, the $4^4$ square-net is a more densely packed configuration where the bond distance between neighboring atoms is $\frac{\sqrt{2}}{2}a$, resulting in twice the atomic density compared to a simple square lattice[11]. This compact arrangement enhances the overlap of in-plane $p_x$ and $p_y$ orbitals, leading to the formation of linearly dispersing Dirac crossings in the band structure (Fig. 5b)[11,211]. Upon introducing SOC and magnetic interactions, many of these nodal lines gap out, giving rise to topological band inversions, massive Dirac fermions, or large Berry curvature hot spots.[210,212-215]. Several families of square-lattice-based compounds have gained prominence in the study of topological quantum materials, particularly structure types like PbFCl (e.g., ZrSiS), *P*4/*nmm* (e.g., $A$MnBi$_2$), and ThCr$_2$Si$_2$ (e.g., LaMn$_2$Ge$_2$) are excellent examples[34,35,215]. Many of these allow chemical substitution, external pressure, or strain tuning, and most importantly for experimental work, high-quality single crystal growth via flux or melt techniques. These materials typically contain layers of *p*-block elements arranged in planar square-nets, where orbital overlap and symmetry constraints give rise to topologically nontrivial electronic structure. Importantly, many of these materials are stable in layered structures that can be synthesized as high-quality single crystals, enabling experimental verification of their topological properties.

### 3.1 AMX$_2$

The general chemical formula $AMX_2$ (where $A$ = Ca, Sr, Ba, Eu, Yb; $M$ = Mn, Zn, Cu, Ag, Au; and $X$ = Sb, Bi, As) encompasses a wide range of layered square-net compounds known for their unique interplay between magnetism and topological electronic states[208,212-214,216-222]. These materials typically crystallize in the tetragonal structure with space groups either *I*4/*mmm* (CaSmP$_2$-type) or *P*4/*mmm* (HfCuSi$_2$-type), featuring planar $4^4$ square-net of transition metal $M$ and pnictogen $X$ atoms and in turn these layers are separated by alkaline-earth or rare-earth $A$[11]. A notable exception is SrMnSb$_2$ and the related materials, which adopt an orthorhombic structure due to lattice distortion, yet retain a Sb square lattice, but it is distorted[208,223]. The square-nets of Bi or Sb are derived by $p_x$ and $p_y$ orbitals, giving rise to highly anisotropic, linearly dispersive Dirac-like states near the $E_F$. Among the most studied members, SrMnBi$_2$ exemplifies the characteristic Dirac semimetal behavior of this family. The research on topological square-net materials is in fact started with the discovery of anisotropic Dirac fermions in SrMnBi$_2$, which crystallizes in *I*4/*mmm*[216]. Its Bi square-net forms highly dispersive Dirac bands, while the Mn atoms exhibit G-type antiferromagnetic order below $T_N$ ~ 290 K, where ferromagnetic layers stack antiferromagnetically along the *c*-axis. This broken time-reversal symmetry, combined with SOC from the heavy Bi atoms, modifies the Dirac states and induces Berry curvature, allowing signatures of non-trivial topology in transport experiments. Another notable example is CaMnBi$_2$, which crystallizes in space group *P*4/*mmm*. It shares similar physics but displays C-type AFM with a different stacking of layers and slightly modified Dirac dispersion, exhibiting quantum magnetotransport response[223,224]. Surprisingly, the isostructural Zn-materials *A*ZnBi$_2$ (A = Ca, Sr, Ba) seem to have no Dirac dispersions[225].

Furthermore, *A*MnBi$_2$ provides an ideal platform for realizing the interplay of Dirac fermions and ordered magnetic moments[226]. This is because Bi square-net conducting layers host quasi-Dirac fermions and the insulating magnetic layers of Mn-Bi and *A* ions are spatially separated, where the different magnetic layers can be constructed while keeping Dirac like band structure. In EuMnBi$_2$, the Eu$^{2+}$ sublattice orders antiferromagnetically below ~22 K and couples to the Bi square-net, enabling the realization of a quantized Hall effect in bulk (Fig. 5c)[213]. This field-induced behavior arises due to the confinement of Dirac fermions in 2D layers under magnetic order and highlights the ability to tune topological phases via magnetic exchange fields. In YbMnBi$_2$, a canted antiferromagnetic structure breaks time-reversal symmetry, enabling the realization of a type-II Weyl semimetal state characterized by finite Berry curvature[214,226]. The presence of heavy Bi atoms introduces strong SOC, which, combined with magnetic symmetry breaking, enhances topological transport responses. Notably, the material exhibits a large anomalous Nernst conductivity ~10 A m$^{-1}$ K$^{-1}$ (Fig. 5d), which is significantly exceeding that of typical ferromagnets and rivalling those found in noncollinear antiferromagnets, despite its layered structure[214]. This highlights the effectiveness of square-net systems in hosting high-performance topological thermoelectric effects. The *AMX*$_2$ materials serve as an exemplary class where square-net-driven Dirac bands coexist and interact with magnetically ordered backgrounds, creating fertile ground for discovering magnetically tunable topological semimetals, quantum Hall systems, and Berry-curvature-driven transport in low-symmetry and low-dimensional settings. This indicates the wide variety of functionalities in these square-net materials.

**3.2 MXX′**

The family of compounds *MXX′* (*M* = Zr, Hf, rare-earth; *X* = Si, Ge, Sn, Sb; *X′* = S, Se, Te) demonstrates exotic nontrivial topological properties[11,35,227-231]. These compounds crystallize in a tetragonal PbFCl-type structure (space group: *P*4/*nmm*), where *X* atoms form the square-net responsible for linearly dispersing bands near the $E_F$. In the absence of rare-earths, these materials exhibit large energy ranges of linear dispersion, up to 2 eV in some cases, and are notable for their relatively clean electronic structures with minimal interference from trivial bands. Their nonsymmorphic symmetry plays a crucial role in protecting band crossings along specific directions (notably *X–R* and *M–A*) in the BZ (Fig. 5e)[35]. ARPES experiments confirm the presence of such line nodes, and these nodes are robust against SOC, especially in ZrSiS, where SOC is weaker. One of the most compelling topological features of ZrSiS is the presence of the Fermi surfaces enclosing Dirac line nodes. The associated Berry phase acquired by electrons circling these nodes contributes to quantum oscillations and magnetotransport signatures, including nontrivial Berry curvature effects and extreme MR. In addition to ZrSiS, its sister materials have also been explored for their potential support to Dirac nodal line electronic structures. In HfSiS, the stronger SOC causes more visible band splitting (Fig. 5e) and modifies the surface state structure, but crucially, the Dirac line nodes remain protected due to nonsymmorphic symmetry, leading to anisotropic magnetotransport,[227,232]. This is one of the examples wherein the effect of SOC, symmetry, and topology can be observed in electrical transport. The difference in SOC strength between Hf and Zr systems allows for a comparative analysis of topological resilience and surface state evolution. It is interesting to point out here that ZrSiTe is the one of the best material to exhibit the Dirac crossing at the $E_F$[233].

The magnetic counterpart of the *MXX′* series is derived when *M* is replaced by magnetic *R* elements. There is an increasing interest in exploring 4f-electron-based systems due to their strong electronic correlations and complex magnetic interactions. The *R*SbTe family serves as a model example in this context, wherein Sb atoms form a square-net[228-231]. For instance, CeSbTe undergoes antiferromagnetic ordering below $T_N$ = 2.7 K, where $Ce^{3+}$ moments align ferromagnetically within the *ab*-plane but stack along the *c*-axis in a nontrivial up-down-down-up arrangement, effectively doubling the magnetic unit cell along the *c*-axis[228]. This complex magnetic structure enables the realization of multiple topological phases, including field-tunable Weyl and Dirac states, depending on the orientation and magnitude of the applied magnetic field. Different magnetic configurations correspond to distinct magnetic space groups, resulting in varied symmetry-protected band crossings. In particular, the antiferromagnetic state hosts an eightfold degenerate point, potentially linked to novel fermionic excitations, making $R$SbTe a versatile platform for exploring magnetism-driven topological phase transitions[11].

Beyond this series, MnAlGe crystallizes in the same tetragonal PbFCl-type structure and the same space group. The structure is fundamentally a layered derivative of the cubic Heusler ($Cu_2MnAl$-type, $Fm$-$3m$) structure, wherein Mn atoms form square-net ($4^4$) layers separated by nonmagnetic Al–Ge spacer layers[234]. Unlike other related Mn-based compounds from this family, which are semiconducting and antiferromagnetic, MnAlGe is metallic and ferromagnetic with a high Curie temperature of $T_C \sim 503$ K, due to the presence of an extra valence electron. MnAlGe hosts a topological nodal line semimetallic state, with multiple nodal lines near the $E_F$. Upon inclusion of SOC results in the emergence of significant 2D Berry curvature, thereby giving rise to a large AHC of 700 $\Omega^{-1}$cm$^{-1}$ at 2 K, as shown in Fig. 5h. This arises due to SOC-induced gaps in nodal lines, which are formed by Mn-d orbitals within the square-net layers. Furthermore, ARPES experiments confirmed the presence of nodal-line dispersions just below the $E_F$, solidifying its topological character (Fig. 5i). Notably, the AHC per magnetic Mn layer approaches the quantum of conductance ($e^2/h$), which is attributed to the out-of-plane ferromagnetic order and the topologically nontrivial band structure of the Mn square-nets. This positions MnAlGe as a rare example of a ferromagnetic nodal-line semimetal with quantized Berry curvature-driven transport, combining layered structure, high-$T_C$ magnetism, and topological phenomena.

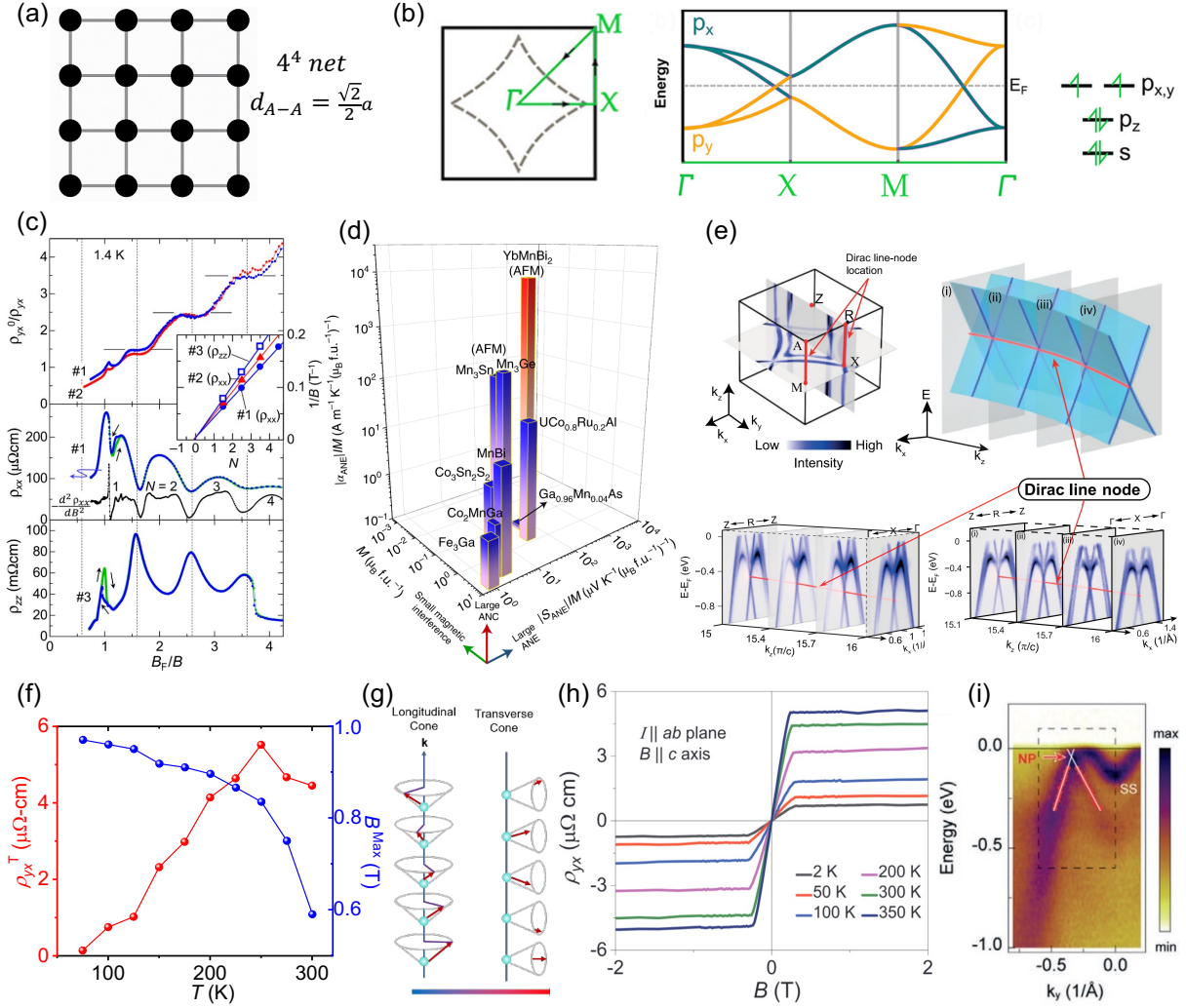

**Fig. 5 Square-net atomic arrangement and its related key phenomena**. **a** Representation of $4^4$ square-net of atoms and the corresponding Brillouin zone. **b** Electronic band structure resulting from a square-net of $p_x$ and $p_y$ orbitals, adapted from Ref.[222]. **c** Bulk quantum Hall effect in EuMnBi$_2$, adapted from Ref.[213]. **d** Anomalous Nernst effect in YbMnBi$_2$ compared with other state-of-the-art materials, adapted from Ref.[214]. **e** ARPES measurements showing the presence of Dirac nodal lines in ZrSiS (left) and HfSiS (right), adapted from Ref.[232]. **f** Temperature and magnetic field dependent topological Hall effect and **g** Magnetic structure evolution from longitudinal cone to transverse cone under magnetic field of LaMn$_2$Ge$_2$, adapted from Ref.[215]. **h** AHC and **i** Band dispersion measure along $\Gamma$–$X$ in MnAlGe. The red arrow marks the band crossing nodal point, adapted from Ref.[234]

### 3.3 RMn$_2$Ge$_2$

$R$Mn$_2$Ge$_2$ series is represented by the ThCr$_2$Si$_2$-type (space group $I4/mmm$), where $R$ is a rare-earth, consists of layered intermetallic compounds where the magnetic Mn atoms form square lattices separated by rare-earth and Ge layers[215,235-240]. The Mn atoms form the square-net within the ab-plane in this family, which is a primary source of magnetism. The square-net atoms interact with R atoms and directly affects the combining property of magnetism and topology. Among the series, the detailed electrical transport investigations of LaMn$_2$Ge$_2$ exhibit a giant THE of ~4.5 μΩcm even at 300 K, which is a record-high value for any known material with a noncoplanar magnetic structure (Fig. 5f). The neutron diffraction experiments further confirm the incommensurate conical spin structure (Fig. 5g), while ARPES and DFT calculations highlighted band inversions and Berry curvature contributions near $E_F$[215]. The combination of high Curie temperature (~325 K), intrinsic anomalous Hall conductivity, and room-temperature topological effects positions $LaMn_2Ge_2$, from square-net-based materials, as a highly promising platform for spintronic applications. Several other compounds from this family exhibit intrinsic tunable electrical and thermal transport properties originating from the noncoplanar ($SmMn_2Ge_2$), helical magnetic ($CeMn_2Ge_2$), and noncolinear ferromagnetic ($PrMn_2Ge_2$) spin structures[238-240].

### 3.4 AX$_4$

The $AX_4$ compounds ($A$ = Ba, Sr, Eu; $X$ = Al, Ga, In) adopt the $BaAl_4$-type tetragonal structure (space group $I4/mmm$), where $X$ atoms form buckled square-nets sandwiched between $A$ cation layers[210,241-243]. These materials exhibit crystalline symmetry-protected non-trivial topological semimetallic behavior, with 3D Dirac band dispersions and nodal-line features in the absence of SOC. Several quantum transport features are exhibited by this series of compounds, including CDW, high mobility, and high MR. In particular, $SrAl_4$ and $EuAl_4$ develop a charge density CDW phase driven by strong electron-phonon coupling associated with the transverse acoustic phonon mode along the $\Gamma$–$Z$ direction[243]. Similarly, $EuGa_4$ has been identified as a rare magnetic Weyl nodal ring semimetal, in which mirror symmetry protects low-dispersion nodal rings near the $E_F$[210]. Upon the introduction of spin polarization via magnetic ordering and SOC, the initially spinless nodal rings split into spinful rings, and mirror symmetry selects only a subset to survive as Weyl nodal rings. The Ga square-net contributes $p_x/p_y$-derived bands responsible for these nodal features, which exhibit clear Landau quantization and give rise to extremely large, non-saturating MR (exceeding 200,000% at 2 K and 14 T). The directional control of the Eu moments allows symmetry-tuning of these topological bands, demonstrating the unique capacity of magnetic square-net materials to host field-manipulated Weyl nodal structures. The square-net topology and symmetry-protected Dirac features, coupled with lattice instabilities, make this family a rich platform for studying the interplay of topology, lattice dynamics, and many-body quantum interactions.

Table 3 | Property, growth, and challenges of square-net materials

| Compound | Growth method | Key property | Key challenge | Reference |
| --- | --- | --- | --- | --- |
| $AMX_2$ ($A$ = Ca, Sr, Ba, Eu, Yb; $M$ = Mn, Zn, Cu; $X$ = Sb, Bi, As) | Flux growth | Dirac semimetals, large MR, high mobility, magnetic tuning, quantum Hall effect | Control of pnictogen volatility (Bi/Sb), layered cleavage, phase purity, precise Mn stoichiometry, flux remnants | 11,208,212-214,217,218,224 |
| $WXX'$ ($W$ = Zr, Hf, rare-earth; $X$ = Si, Ge, Sn, Sb; $X'$ = S, Se, Te) | Flux growth or CVT | Nodal-line semimetals, large non-saturating MR, surface states | Si volatility, phase competition, air sensitivity, maintaining flat square-nets | 11,35,227,232,233,244 |
| $R$SbTe ($R$ = rare-earth) | Flux growth | Magnetic Dirac/Weyl states, tunable degeneracies, symmetry-driven transitions, CDW | precise control of rare-earth, air sensitive | 228-231 |
| $R$Mn$_2$Ge$_2$ ($R$ = rare-earth) | Flux growth | Noncoplanar magnetism, skyrmions, Dirac dispersions | Mn oxidation, controlling layer uniformity, high-temperature phase formation, competing secondary phases | 215,235-240 |
| $MX_4$ ($M$ = Ba, Sr; $X$ = Al, Ga, In) | Flux growth | Nodal-line ferromagnet, large AHC | Suppressing antisite disorder, stabilizing ferromagnetic phase, maintaining $4^4$ square-net integrity | 210,241,243 |

## 3.5 Crystal growth of square-nets materials

The details of different growth methods have already been discussed in the previous sections of kagome and chiral. Thus, this section will address the growth conditions of a select number of square-net materials. For instance, the crystal growth of the $AMX_2$ family is typically carried out using Bi/Sb self-flux. One of the primary challenges lies in the tendency for Bi/Sb-rich impurity phases (like Bi, Sb, MnBi, and MnSb) to crystallize alongside the desired phase. The presence of residual flux in the form of droplets appears on the surface of the crystals[245]. Successful growth requires careful tuning of the starting stoichiometry, use of a high excess Bi/Sb flux, and decanting at the right temperature (typically ~1000–1100 °C, followed by slow cooling and decanting around 600–700 °C). The plate-like morphology makes the crystals prone to cleavage, but this is also advantageous for transport and ARPES studies. These materials are reactive in air thus, the handling is carried out in an inert gas atmosphere as far as possible. Single crystals of ZrSiS-type $MXX'$ compounds ($M$ = Zr, Hf; $X$ = Si, Ge, Sn; $X'$ = S, Se, Te) are typically grown by CVT using halide transport agents. However, this method proves ineffective for certain members like ZrSnTe, primarily due to competing binary phases and challenges in achieving supersaturation[244]. In the case of ZrSnTe, flux growth using Sn as a self-flux has proven successful. The key lies in optimizing flux ratio, cooling rate, and temperature profile to suppress impurity phases such as SnTe or $ZrTe_2$, which easily form during slow cooling. Fast cooling (~50 °C/hr) from 1000 °C is essential to obtain high-quality single crystals of ZrSnTe, revealing that the desired phase may be metastable, stabilized only under non-equilibrium conditions. This highlights a broader challenge in the *MXX'* family. However, CVT is a standard technique; flux growth becomes crucial for some of the selected compounds, requiring careful control of thermodynamics and kinetics to achieve clean, phase-pure topological crystals. Single crystals of $RMn_2Ge_2$ (*R* = rare-earth or alkaline-earth elements) have been successfully grown using indium (In) flux, which offers a cleaner growth environment compared to self-flux. The melting point of In is nearly six times lower than Ge and no binary phases of Ge-In exist. Thus, In serves as a low-melting, chemically inert medium that allows for slow and controlled crystallization, promoting high-quality plate-like crystals with fewer inclusions. For instance, high-purity elemental *R* (99.9%), Mn, Ge, and In are mixed with a 1:2:2:20 molar ratio in an alumina crucible. The sealed crucible is heated to 1150 °C K, and then cooled slowly to 700 °C at a rate of 3 °C/h. After the growth, excess indium is typically removed by centrifugation at high temperatures, followed by selective etching in HCl or dilute $HNO_3$ to clean residual flux. This method yields phase-pure $RMn_2Ge_2$ crystals with well-developed facets, suitable for detailed magnetic and transport property studies.

**Table 4 | Combined comparison of kagome, chiral and square-net quantum materials**

| Feature | Kagome material | Chiral material | Square-net material |
| --- | --- | --- | --- |
| Lattice geometry | 2D network of corner-sharing triangles | Non-centrosymmetric 3D (helical chains, asymmetric layers) | Planar $4^4$ square-nets of p-block atoms |
| Key symmetries | Geometric frustration, $C_{6v}$ symmetry | Broken inversion/mirror; enantiomorphic | Nonsymmorphic symmetries (glide mirrors, screw axes) |
| Topological fermions | Dirac, flat bands, Weyl (via magnetism) | Kramers-Weyl, multifold fermions | Dirac, nodal-line, Weyl (with magnetism) |
| Signature phenomena | CDW, pair density waves, AHE, topological superconductivity | Skyrmions, chiral soliton lattices, nonlinear Hall effects | Drumhead surface states, tunable Dirac nodes, Berry-curvature-driven transport |
| Magnetism | Collinear and non-collinear AFM/FM; tunable in $RM_6X_6$ | DMI-driven helimagnetism; spin textures coupled to chirality | Magnetic ions induce Weyl physics; AHE, ANE |
| Notable materials | $Mn_3Sn$, $CsV_3Sb_5$, $Co_3Sn_2S_2$, $TbMn_6Sn_6$ | MnSi, CoSi, PdGa, $Cr_{1/3}NbS_2$, $RPt_2B$ | ZrSiS, $EuMnBi_2$, $SrMnSb_2$, $YbMnBi_2$ |

| Synthesis methods | Flux (Sn, Sb, Bi), Bridgman, CVT | CZ, FZ, Bridgman, CVT | Flux, CVT, Bridgman |

**Challenges and outlook**

Topological quantum materials based on kagome, chiral and square-net frameworks represent a frontier of condensed matter research, offering a rich variety of physical properties. Nevertheless, several challenges hinder their exploration and synthesis of high-quality single crystals remains difficult, particularly with regard to maintaining stoichiometry. The various phenomena, which are not yet fully understood, require further theoretical models and appropriate experimental tools to prove them. Despite the challenges encountered, the future outlook is optimistic.

Kagome compounds with a high level of degeneracy are ideal for achieving strongly correlated electronic states such as superconductivity, fractional quantum anomalous Hall effect, excitonic insulating state and magnetism. Several kagome structural families have been observed to date, ranging from non-collinear magnets of $M_mX_n$ series, intertwined orders of $A$V$_3$Sb$_5$ to tunable magnetism in $R$M$_6$X$_6$. This class of materials has been found to host a remarkable spectrum of phenomena, including Dirac fermions, FB, vHS, CDW, AHE, and unconventional superconductivity. A considerable number of these states have been thoroughly discussed and reported, while the fundamental explanations of some of them remain incomplete. The question arises as to how AHE-like signal can exist in the absence of magnetism. It is of interest to determine whether distorted kagome materials will access chiral phonons or host new classes of topological excitations. This series has not yet been thoroughly investigated, and thus, fundamental inquiries persist. For instance, it is as yet unclear whether distortion exerts control over the interplay between topology and frustration. A number of module systems have been proposed for field-free switching applications, which is a promising development. These systems are predicated on non-collinear spin textures, exchange bias effects, and large spin Hall angles[246,247]. Electrical spin-orbit torque switching based on Mn$_3$Sn heterostructures (e.g., Mn$_3$Sn/W) has been demonstrated to reliably generate binary memory states (0/1) via Hall voltage polarity[248]. Furthermore, Kagome materials promise for use in twistronics through the formation of Moiré superlattices and to achieve pressure- or strain-driven quantum critical points.

Chirality is a crucial factor in advanced-level applications in pharmaceuticals because many drugs exist in chiral forms. Often, one enantiomer is more effective or safer than the other.

Significant progress has been made in the development of inorganic chiral quantum materials. The mechanisms underlying chiral-induced spin selectivity remain to be fully elucidated, particularly in systems exhibiting strong electron correlations and SOC. The relationship between chirality, lattice dynamics, phonon modes, and their influence on transport properties and phase transitions is unclear. In the context of chiral superconductors, the influence of structural chirality on the phenomenon of superconductivity remains to be understood. The question of whether chirality enhances pairing mechanisms or leads to unconventional superconducting states with broken time-reversal symmetry remains a subject of debate. Furthermore, the occurrence of superconductivity in polar chiral materials offers a means to explore the interplay between structural chirality, polarity, and unconventional superconducting pairing mechanisms[249]. The incorporation of chiral materials into heterostructures has the potential to enhance the functionality of spintronics. The viability of these applications is contingent upon the complete control of chirality during the synthesis of chiral materials. The homochiral crystals can be successfully grown using the LFZ and Czochralski methods, but both these techniques rely on a seed crystal with the desired handedness. Once growth begins, the resulting crystal maintains a single handedness and it is no longer possible to control chirality. Therefore, it would be valuable to explore whether chirality can be actively controlled or even reversed during the growth process. Given the sensitivity of chiral materials to circularly polarized light, magnetic or electric fields, it is reasonable to hypothesize that these external stimuli could be used to influence crystal handedness. A substantial OAM of chiral materials has the potential for efficient current generation and manipulation without SOC, thereby expanding the scope of orbitronics.

Several promising directions can propel the field of square-net quantum materials. Their characteristic nonsymmorphic symmetries support Dirac and nodal-line fermions, while the inclusion of SOC and magnetic symmetry breaking can give rise to Weyl nodes and large Berry curvature effects, as observed through anomalous Hall and Nernst responses. Despite these advances, key questions remain open. For instance, the stability of topological surface states in the presence of disorder, the influence of phonons and CDWs on nodal-line instabilities, and the possibility of correlation-driven topological superconductivity are yet to be fully understood. Additionally, the role of electron-phonon coupling and lattice softness in facilitating emergent orders near van Hove singularities or flat bands remains a fertile ground for exploration. One exciting frontier lies in the realization of chiral phonons, defined as phonon modes characterized by finite angular momentum. These phonons emerge in response to the breakdown of inversion or mirror symmetries, through processes such as buckling, magnetic order, or interlayer coupling, enabling topological phonon bands or the phonon Hall effect. In parallel, advances in artificial lattice engineering, such as optical lattices and moiré superstructures, provide unprecedented control over hopping parameters, symmetry, and dimensionality. This facilitates the design of Chern insulators, flat-band systems, and Floquet topological phases in square-net geometries. The expansion of material library through targeted synthesis and symmetry-guided design, particularly with elements exhibiting strong SOC or intrinsic magnetism, will be essential to uncover new regimes where topological and collective excitations coexist and interact. These directions will not only serve to further refine our comprehension of square-net systems, but they may also lay the foundational groundwork for the development of prospective quantum technologies in the future.

**Acknowledgements**

The work was financially supported by the Deutsche Forschungsgemeinschaft (DFG, German Research Foundation) through SFB 1143 (project ID 24731007), QUAST (project ID FOR 5249), the Würzburg-Dresden Cluster of Excellence on Complexity and Topology in Quantum Matter—ct.qmat (EXC 2147, project ID 390858490), and EXQIRAL (No. 101131579).